\documentclass[prL,superscriptaddress,reprint]{revtex4-2}

\usepackage{amsmath}
\usepackage{graphicx}
\usepackage{bm}
\usepackage{MnSymbol}
\usepackage{eufrak}
\usepackage{amsfonts} 
\usepackage{braket}
\usepackage{hyperref}

\usepackage{xcolor}
\newcommand{\rev}[1]{{#1}}

\begin{document}

\title{Topological quantum synchronization of fractionalized spins}

\author{Christopher W. W\"achtler}
\email{cwwaechtler@berkeley.edu}
\affiliation{Department of Physics, University of California, Berkeley, California 94720, USA}
\author{Joel E. Moore}
\affiliation{Department of Physics, University of California, Berkeley, California 94720, USA}
\affiliation{Materials Sciences Division, Lawrence Berkeley National Laboratory, Berkeley, California 94720, USA}

\date{\today}

\begin{abstract}
The gapped symmetric phase of the Affleck-Kennedy-Lieb-Tasaki (AKLT) model exhibits fractionalized spins at the ends of an open chain. We show that breaking SU(2) symmetry and applying a global spin-lowering dissipator achieves synchronization of these fractionalized spins. Additional local dissipators ensure convergence to the ground state manifold. In order to understand which aspects of this synchronization are robust within the entire Haldane-gap phase, we reduce the biquadratic term which eliminates the need for an external field but destabilizes synchronization. Within the ground state subspace, stability is regained using only the global lowering dissipator. These results demonstrate that fractionalized degrees of freedom can be synchronized in extended systems with a significant degree of robustness arising from topological protection. \rev{A direct consequence is that permutation symmetries are not required for the dynamics to be synchronized, representing a clear advantage of topological synchronization compared to synchronization induced by permutation symmetries.}
\end{abstract}

\maketitle

\emph{Introduction.---}From neuroscience to chemical reactions, synchronization emerges in an impressively vast variety of seemingly unrelated systems \cite{pikovsky2003synchronization, StrogatzBook2018,Strogatz1993, Rosenblum2003, arenas2008synchronization} and despite its long history continues to be crucial for the development of modern technology \cite{thornburg1997chaos, lynch1995mode,cawthorne1999synchronized, fazio2001quantum, slavin2009spin, Nishikawa_2015, bellamy1995digital, narula2018requirements}. In the past decade, the concept of synchronization has been generalized to the quantum regime with studies ranging from classically inspired systems like nonlinear oscillators \cite{Lee2013, PhysRevLett.112.094102,dutta2019critical, Walter2015, bastidas2015, PhysRevE.96.052210, 2013-Fazio-PRL, PhysRevA.106.012422, ameri2015mutual, sonar2018squeezing, es2020synchronization,lorch2016genuine, Davis2018, amitai2018quantum, mok2020synchronization, PhysRevResearch.5.023021, delmonte2023quantum, shen2023enhancing} to systems without any classical counterpart like spins \cite{PhysRevLett.121.053601, PhysRevLett.121.063601, daniel2023geometric, schmolke2022noise, schmolke2023measurement}. Mutual synchronization and forced synchronization have been examined with surprising effects that are absent in the classical regime, such as for example the phenomenon of synchronization blockade of two identical systems \cite{PhysRevLett.118.243602, tan2022half}, which has recently also been verified experimentally \cite{PhysRevResearch.2.023026}. Promising applications of quantum synchronization range from quantum information \cite{pljonkin2021vulnerability, pljonkin2017synchronization, liu2019secure, agnesi2020simple} to quantum thermodynamics \cite{jaseem2020quantum, murtadho2023cooperation, murtadho2023synchronization}.

One approach to synchronization -- followed in particular in the study of quantum many-body systems -- is in terms of persistently oscillating eigenmodes of time-independent quantum master equations \cite{schmolke2022noise, buca2022algebraic, Tindall2020}. The existence of such eigenmodes is intimately related to dynamical symmetries \cite{buvca2019non, buca2022algebraic, Tindall2020}, which together with permutation symmetries allow for synchronized dynamics of local observables. An illustrative example investigated in Ref.~\cite{buca2022algebraic} is a 3-site Hamiltonian which non-trivially couples three spin-1/2 particles, is reflection symmetric about the central site, and conserves  total magnetization. Dissipation acting locally on the central spin forces it to be in the spin-down state. As a consequence, there are two steady states of the master equation, where the two remaining spins are both spin-down  or form a singlet. These two pure states form a decoherence-free subspace \cite{lidar1998decoherence} such that coherent oscillations between these two states are possible even in this dissipative setup. Starting in an initial state that has non-vanishing overlap with both the singlet and the both-down state, results (after a short transient time) in 
perfectly anti-synchronized oscillations  of the local transverse spin of the non-central sites $2$ and $3$, i.e., $\braket{\sigma_2^\mathrm{x}(t)} = -\braket{\sigma_3^\mathrm{x}(t)}$. They are anti-synchronized because the singlet state is anti-symmetric upon reflection while the spin-down state is symmetric. In the  corresponding Bloch sphere representation, the central spin rapidly decays to the south pole, while the other two spins reach the same limit cycle within the Bloch sphere (parallel to the x-y-plane) which they orbit perfectly out of phase. 

In this Letter we investigate whether a similar strategy can be exploited to synchronize the fractionalized spin-1/2 degrees of freedom localized at the open ends of a spin-1 AKLT chain \cite{PhysRevLett.59.799, haldane1983continuum, PhysRevB.81.064439}. By applying dissipation that acts globally on all sites, we show that lifting the ground state degeneracy through a small external magnetic field leads to stable synchronization of the fractionalized spins. In that case, local spin-1 observables at the ends are perfectly anti-synchronized with amplitudes that reflect the topological edge states, i.e. they are exponentially localized at the boundaries. In addition we show that quasi-local dissipators acting on two neighboring sites that dissipate the energetically lowest state of the total spin $S=2$ subspace are sufficient to depopulate the whole excited subspace and remove unwanted additional oscillations. The observed synchronization is of topological nature as the underlying mechanism relies on the fractionalization of the spin degrees of freedom and is thus \rev{topologically} protected \rev{by the $\mathbb Z_2\times \mathbb Z_2$ symmetry of the AKLT states}. \rev{Consequently, the observed synchronization is robust to perturbations that break the inversion (or permutation) symmetry, which can induced additional dissipation and eliminate long-lasting synchronized dynamics.} Lastly, we show that if the fractionalized spins are allowed to interact by decreasing the biquadratic term of the AKLT Hamiltonian, stable synchronization within the ground state manifold \rev{of the Haldane-gap phase} is still possible even without external magnetic field if one only considers a global spin lowering dissipator. This demonstrates that the dynamic response depends on the microscopic details of systems even though they belong to the same symmetry protected topological phase.

\emph{Synchronization model.---}We consider the open spin-1 AKLT chain of size $N$ with an additional external magnetic field $B$ yielding the Hamiltonian
\begin{equation}
\label{eq:Hamiltonian_AKLT}
    H = \sum\limits_{j=1}^{N-1} \left[\frac{1}{2}\vec{S}_j\cdot \vec{S}_{j+1} + \frac{1}{6}\left(\vec S_{j}\cdot \vec S_{j+1}\right)^2 + \frac{1}{3}\right] +\frac{B}{N} S^\mathrm{z},
\end{equation}
where $S^\mathrm{z} = \sum_{j=1}^N  S_j^\mathrm{z}$ is the total magnetization. For sufficiently small values of $B$, the Hamiltonian remains gapped even if the chain size is increased, yet breaks  SU(2) symmetry (which will become important for synchronization as we explain later). For $B=0$ the ground state is fourfold degenerate as a consequence of effective spin-1/2 degrees of freedom that are localized at both ends of the chain.  The ground states of (\ref{eq:Hamiltonian_AKLT}) can be constructed explicitly, e.g., in terms of Schwinger bosons \cite{PhysRevB.98.235155} or matrix product states \cite{schollwock2011density, 10.5555/2011832.2011833, orus2014practical}. As the spin-1/2 degrees of freedom at the ends are exactly decoupled, there are three ground states with total spin $S=1$, where the two dangling spin-1/2's form a triplet state with $S_\mathrm{z} = 1,0,-1$ and one with $S = 0$, where the dangling spin-1/2's form a singlet with $S_\mathrm{z}=0$. Thus, we may label the ground states accordingly as $\ket{G_{1,1}}$, $\ket{G_{1,0}}$, $\ket{G_{1,-1}}$ and $\ket{G_{0,0}}$. Note, that while a finite value of $B$ partially lifts the ground state degeneracy, the corresponding manifold is still spanned by $\left\{\ket{G_{S,S_\mathrm{z}}}\right\}$ as the total magnetization is preserved;  $\left[H,S^\mathrm{z}\right]=0$. 

Synchronization is inherently connected to open system dynamics because it requires dissipation in order to reduce all potential dynamics to only the desired, synchronized ones. To this end, we describe the system by a time dependent density operator $\varrho(t)$ acting on the Hilbert space of the system $\mathcal H$. We consider Markovian dynamics such that the evolution may be described via a Lindblad master equation \cite{BreuerPetruccioneBook2002, SchallerBook2014},
\begin{equation}
\label{eq:Lindblad}
\dot\varrho = -\mathrm i \left[H,\varrho\right] + \sum\limits_\mu \left(2 L_\mu \varrho L_\mu^\dagger - \left\{ L_\mu^\dagger L_\mu, \varrho\right\}\right)= \mathcal L\left[\varrho\right],
\end{equation}
where $L_\mu$ denotes (for now unspecified) Lindblad operators. The Liouvillian superoperator $\mathcal L$ is the generator of a smooth, time-homogeneous, completely positive and trace-preserving (CPTP) map (or quantum channel), which obeys the semi-group property. 
The system dynamics described by Eq.~(\ref{eq:Lindblad}) is guaranteed to have at least one steady state $\varrho_{\mathrm{ss}}$ such that $\mathcal L\left[\varrho_\mathrm{ss}\right] = 0$ \cite{BreuerPetruccioneBook2002}. 

A sufficient and necessary condition for the existence of an eigenstate  $\varrho = A \varrho_\mathrm{ss}$ of $\mathcal L$ with purely imaginary eigenvalue $\lambda = -\mathrm i \omega$, i.e., $\mathcal L[\varrho]=-\mathrm i \omega \varrho$ with $\omega \in \mathbb R$, is given by \cite{buca2022algebraic}
\begin{align}
\left[L_\mu ,A\right]\varrho_\mathrm{ss} &= 0, \label{eq:Theorem1a}\\
\left(-\mathrm i \left[H,A\right] - \sum\limits_\mu \left[L_\mu^\dagger, A\right]L_\mu \right)\varrho_\mathrm{ss} &= -\mathrm i \omega A \varrho_\mathrm{ss}.\label{eq:Theorem1b}
\end{align} 
While Eqs.~(\ref{eq:Theorem1a}) and (\ref{eq:Theorem1b}) guarantee the existence of persistent oscillations in the long time limit, one usually demands another condition for (anti-)synchronization \cite{buca2022algebraic}. Let $P_{jk}$  be an operator that exchanges subsystem $j$ with $k$ and let $\mathcal P_{jk}[x] = P_{jk}xP_{jk}$. Then, if $P_{jk}$ is a weak symmetry of the Liouvillian, i.e $\left[\mathcal L, \mathcal P_{jk}\right] = 0$, and (anti-)commutes with the operator A, $P_{jk} A P_{jk}=\pm A$, then we find \emph{stable} synchronization ($+$) or anti-synchronization ($-$) of the two local operators $O_j$ and $O_k$ if $\text{Tr}\left[O_j A \varrho_\mathrm{ss}\right]\neq 0$ and conditions (\ref{eq:Theorem1a}) and (\ref{eq:Theorem1b}) are fulfilled, that is after some transient time  $\tau$ 
\begin{equation}
\label{eq:SynchObs}
\braket{O_j(t)} = \pm \braket{O_k(t)}~\forall t\geq \tau
\end{equation}
up to exponentially small corrections. In the example referred to in the introduction the local transverse spin of the non-central sites $2$ and $3$ will be perfectly anti-synchronized, i.e., $\braket{\sigma_2^\mathrm{x}(t)} = -\braket{\sigma_3^\mathrm{x}(t)} \propto \cos(\omega t)$, where the oscillation frequency $\omega$ depends on the specific choice of Hamiltonian \cite{buca2022algebraic}. As we discuss later, the necessity of permutation symmetry is omitted for topological synchronization in the AKLT chain as long as the $\mathbb Z_2\times \mathbb Z_2$ is preserved.

\begin{figure*}
    \centering
    \includegraphics[width=0.95\textwidth]{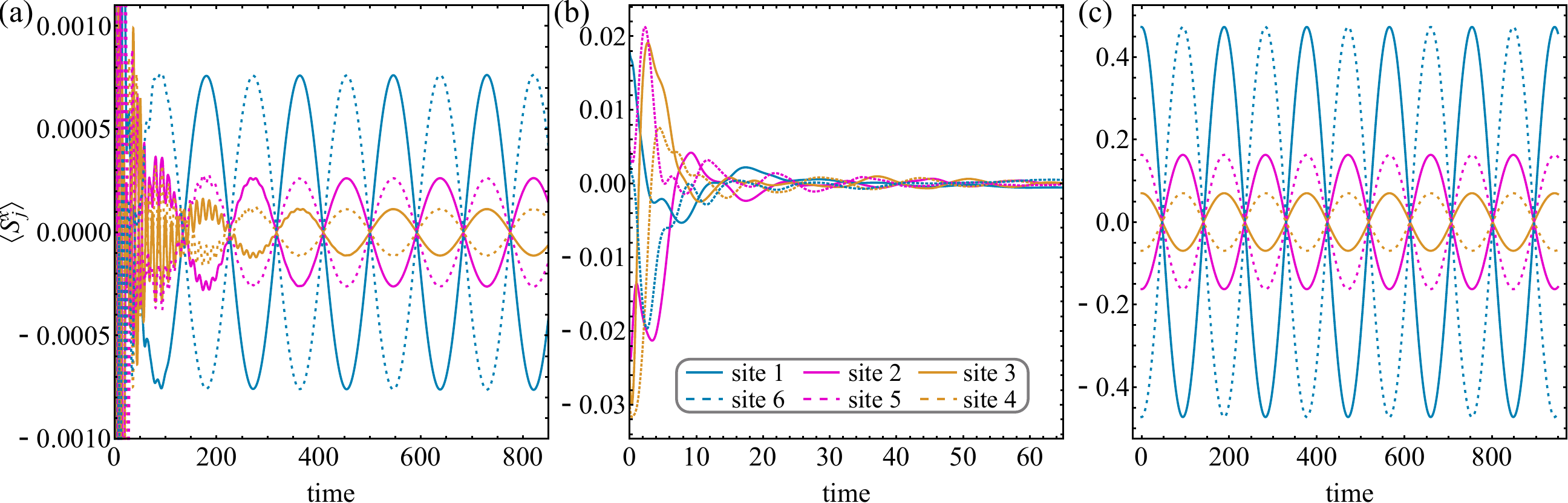}
    \caption{Evolution of the local transverse spin $\braket{S_j^\mathrm{x}}$ of the synchronized AKLT model for an open chain of length $N=6$ (sites $j=1,2,3$ in solid lines, sites $j=4,5,6$ in dashed lines). (a) Starting from \rev{a} random \rev{pure state, i.e., a vector of dimension $3^6$ with random complex amplitudes (we also investigated random mixed states with equivalent results)}, the two halves of the chain are perfectly anti-synchronized with each other after a transient time because the dynamical symmetry operator $A=\ket{G_{1,-1}}\bra{G_0,0}$ is anti-symmetric upon inversion of the chain. The (anti-)synchronized amplitudes after the transient time decay exponentially into the bulk. (b) Same plot as in (a) but focusing on the early time dynamics: The random initial conditions result in transient random spin dynamics. (c) The balanced superposition of $\ket{G_{0,0}}$ and $\ket{G_{1,-1}}$ as initial state is immune to dissipation and maximizes the observed (anti-)synchronization amplitudes. The oscillation frequency is $\omega=B/N$. Parameters: $B=0.2$, $\gamma = \kappa = 0.2$.}
    \label{fig:AKLT}
\end{figure*}

In the following we first focus on the ground state manifold and show how a single, globally acting dissipator $L_\mathrm{G}$ leads to the fulfilment of conditions~(\ref{eq:Theorem1a})-(\ref{eq:SynchObs}) within the ground state manifold and thus to stable synchronization. In a second step we will then show that additional, locally acting dissipators force the dynamics into the ground state manifold. 

In order to find adequate dissipators such that Eqs.~(\ref{eq:Theorem1a}) and (\ref{eq:Theorem1b}) are fulfilled, we utilize the fractionalized spins of the AKLT ground states: Since the triplet and singlet states have different respective total spin $S=1$ and $S=0$, a global lowering operator $S^-=\sum_{j=1}^N S_j^-$ leaves the singlet state $\ket{G_{0,0}}$ invariant while lowering the magnetization $S_z$ of the triplet states. Repeated application of $S^-$ will then force the population into the state with the lowest weight, i.e., $\ket{G_{1,-1}}$, which is also invariant upon acting with $S^-$. Hence, a globally acting Lindblad dissipator $L_\mathrm{G}=\sqrt{\gamma}S^-$ with dissipation rate $\gamma$, establishes two steady states of the master Eq.~(\ref{eq:Lindblad}) given by the pure states $\varrho_{0} = \ket{G_{0,0}}\bra{G_{0,0}}$ and $\varrho_{1} = \ket{G_{1,-1}}\bra{G_{1,-1}}$. 
Together with the operator $A = \ket{G_{1,-1}}\bra{G_{0,0}}$ conditions~(\ref{eq:Theorem1a}) and (\ref{eq:Theorem1b}) are fulfilled; in particular it holds that
\begin{align}
    \mathcal L\left[\varrho_{10}=A\varrho_{0}\right] = \mathrm i \frac{B}{N}\varrho_{10},~  \mathcal L \left[\varrho_{01}=\varrho_{0}A^\dagger\right]= -\mathrm i \frac{B}{N} \varrho_{01}.
\end{align}
Note, that $\varrho_{1} = A\varrho_{0} A^\dagger$. We now also recognize that lifting the ground state degeneracy is necessary to observe synchronization, i.e., without the external magnetic field in Eq.~(\ref{eq:Hamiltonian_AKLT}) the oscillation frequency would be zero. 

\emph{Depopulating the excited states.---}So far we have discussed how synchronization may arise within the ground state manifold with the help of a dissipative channel in terms of $L_\mathrm{G}$. However, in addition the excited states need to be depopulated. This can be done either in a two-step process where one prepares the ground state subspace first using established approaches \cite{kraus2008preparation, zhou2021symmetry, PhysRevLett.95.110503, PhysRevB.91.195143, zhou2021symmetry, PhysRevResearch.5.L022037, kaltenbaek2010optical, PhysRevResearch.5.013190, 10.21468/SciPostPhys.15.4.170, PRXQuantum.4.020315, wang2023dissipative} or via depopulation during the dissipative evolution. As a proof of principle, we here opt for the latter and construct the simplest possible operators by exploiting that the Hamiltonian~(\ref{eq:Hamiltonian_AKLT}) preserves total angular momentum. In particular, for $B=0$, each term in $H$ can be written as $P_{j,j+1}^{(2)}$, where $P_{j,j+1}^{(2)}$ denotes the projector of two spin-1's on sites $j$ and $j+1$ onto total spin-2. Hence, the ground states are reached by driving two adjacent spin-1 particles out of the $S=2$ subspace. The previously introduced  dissipative channel ($L_\mathrm{G}=\sqrt{\gamma}S^-$) forces all population within the $S=2$  subspace to eventually reach the $S_z=-2$ state. Thus, we only need to depopulate these states to dissipatively reach the ground state manifold. An exemplary choice is the Lindblad dissipators $L_{j,j+1}=\sqrt{\kappa}\ket{00}\bra{--}_{j,j+1}$ written in the $S_\mathrm{z}$ basis \{$\ket{+},\ket{0},\ket{-}$\}. 

\emph{Synchronized dynamics.---}
Combining all Lindblad operators, the dissipative evolution of the density matrix which eventually leads to the synchronization of the fractionalized spins is given by
\begin{equation}
    \label{eq:FinalME}
    \dot\varrho = -\mathrm i \left[H,\varrho\right] + \mathcal D\left[L_\mathrm{G}\right]\varrho + \sum\limits_{j=1}^{N-1}\mathcal D\left[L_{j,j+1}\right]\varrho = \mathcal L \left[\varrho\right],
\end{equation}
where $\mathcal D\left[L\right]\varrho = 2L\varrho L^\dagger - \{L^\dagger L,\varrho\}$. Its solution, given that the system is initialized in the state ${\varrho(0)}$, may be expressed using the spectral decomposition of the Liouvillian superoperator as
\begin{equation}
\label{eq:Dynamics}
    \varrho(t) = \sum\limits_k C_{k}\exp\left(\lambda_k t\right)  \varrho_k,
\end{equation}
where $\varrho_k$ is the right eigenstate of  $\mathcal L$ with corresponding eigenvalue $\lambda_k$, i.e., $\mathcal L\left[\varrho_k\right]=\lambda_k \varrho_k$. As $\mathcal L$ is non-Hermitian, the left eigenstates defined by  $\mathcal L^\dagger\left[\sigma_k\right]=\lambda_k^\ast \sigma_k$ may differ from the right ones. However, it holds that $\mathrm{Tr}(\sigma^\dagger_k \varrho_{k'})=\delta_{kk'}$. The constant $C_k$ in Eq.~(\ref{eq:Dynamics}) denotes the overlap of the eigenstates with the initial state $\varrho(0)$, i.e., $C_k=\mathrm{Tr}\left[\sigma_k^\dagger \varrho(0)\right].$ Note that because $\mathcal L$ generates a CPTP map, the eigenvalues $\lambda_k$ can lie only in the left half of the complex plane with $\mathrm{Re}[\lambda_k]\leq 0$, and they always come in pairs, i.e., if $\lambda_k$ is an eigenvalue, so is $\lambda_k^\ast$.  All eigenstates of $\mathcal L$ with negative real part of the corresponding eigenvalues  will experience selective decay, and only the ones which lie on the imaginary axis contribute to the dynamics in the long time limit. 

\begin{figure*}
    \centering
\includegraphics[width=0.95\textwidth]{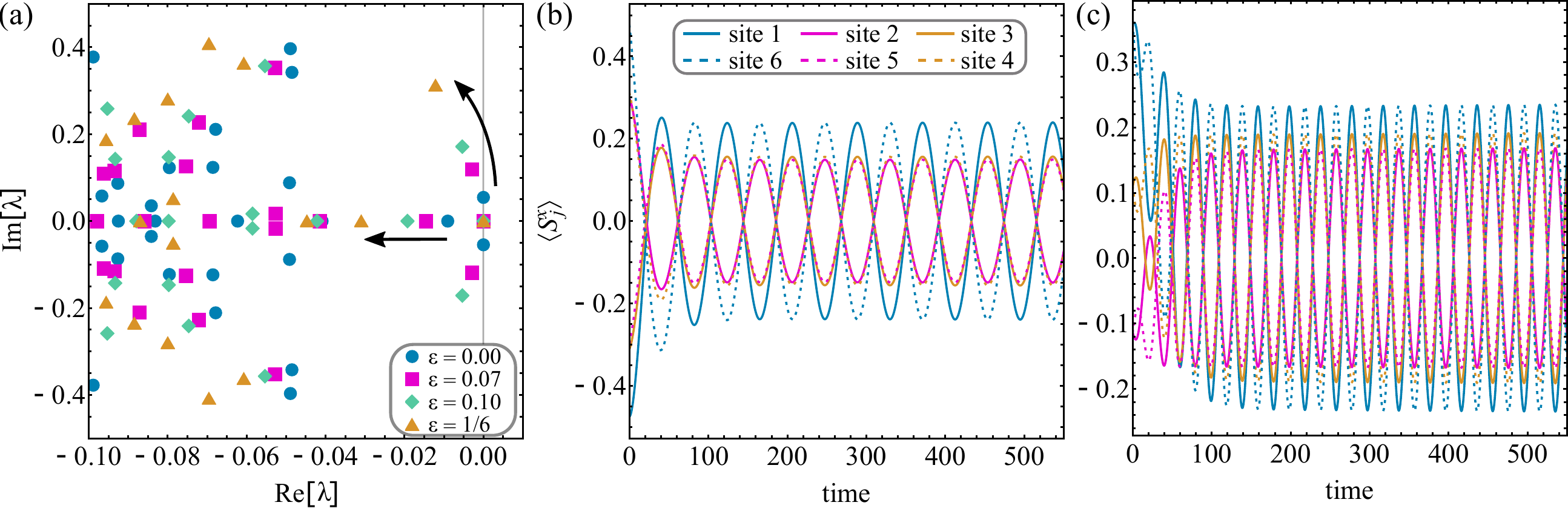}
    \caption{(a) Eigenvalues of the Liouvillian superoperator $\mathcal L$ close to the imaginary axis for  an open chain of length $N=6$ and different values of $\varepsilon$,  considering both global and local dissipators. The purely imaginary eigenvalues for $\varepsilon=0$ move away from the imaginary axis as the biquadratic term is decreased. Simultaneously the oscillation frequency increases resulting in fast but damped (anti-)synchronization. Parameters: $B=0.2$, $\kappa = \gamma = 0.2$. (b, c) Stable synchronization may be recovered for finite values of $\varepsilon$ within the ground state subspace for $B=0$ if one considers only the global dissipator $L_\mathrm{G}$ (i.e. $\kappa=0$). Sites $j=1,2,3$ are in solid lines, sites $j=4,5,6$ in dashed lines. The oscillation frequency in panel (c) with  $\varepsilon=1/6$ is larger compared to panel (b) with $\varepsilon=0.1$ because of the increased gap above the threefold degenerate ground state. The initial state is the infinite temperature state within the ground state manifold. }
    \label{fig:Heisenberg}
\end{figure*}

As discussed previously, the dynamics given by Eq.~(\ref{eq:FinalME}) will eventually terminate in the decoherence-free subspace \cite{lidar1998decoherence} spanned by ${\{\varrho_0},{\varrho_1} \rev{,\varrho_{10}, \varrho_{01}}\}$. Thus, the expectation value of some observable $O$ is given  by
\begin{equation}
\begin{aligned}
\label{eq:ExpectationValue}
    \lim\limits_{t\to\infty}\braket{O}(t) =& C_{0}\mathrm{Tr}\left(O\varrho_0\right) + C_{1}\mathrm{Tr}\left(O\varrho_1\right) \\&+ \left[\mathrm{e}^{\mathrm i B t/N}C_{01}\braket{G_{0,0}|O|G_{1,-1}} + \mathrm{c.c.}\right].
    \end{aligned}
\end{equation}
 Because the subspace is decoherence-free, $C_i =\mathrm{Tr}[\sigma_i^\dagger\varrho(0)]=\mathrm{Tr}[\varrho_i^\dagger\varrho(0)]$. In order to observe stable synchronization, not only does the initial state need to have non-vanishing overlap with the eigenstate ${\varrho_{01}}$, but also that the \rev{observable} is non-zero in that state, i.e., $\mathrm{Tr}(O A\varrho_0) = \braket{G_{0,0}|O|G_{1,-1}}\neq 0$. A suitable choice of local operators that may be used as witnesses of the fractionalized spin synchronization are given by the transverse spin $S_j^\mathrm{x}$, for which the first two terms in Eq.~(\ref{eq:ExpectationValue}) are identical to zero, and only $C_{01} = \braket{G_{0,0}|\varrho(0)|G_{1,-1}}$ and $\braket{G_{1,-1}|S_j^\mathrm{x}|G_{0,0}}$ contribute to the long-time dynamics.

As the dynamical symmetry operator $A=\ket{G_{1,-1}}\bra{G_{0,0}}$ is anti-symmetric upon inversion of the chain, an operator acting locally on site $j$ will be anti-synchronized with the corresponding site at the other end of the chain located at $(N+1)-j$. Figure~\ref{fig:AKLT}(a) and (b) show the time evolution of the transverse spin $\braket{S_j^\mathrm{x}}$ for a chain of length $N=6$ with \rev{a} random \rev{pure state as} initial condition (solid lines correspond to the left half of the chain $j=1,2,3$, dashed lines to the right half $j=4,5,6$). The oscillations are perfectly anti-synchronized upon inversion of the chain. As a consequence of the fractionalized spin, the amplitudes decay exponentially into the bulk. As seen in Fig.~\ref{fig:AKLT}(b) for short times there is no synchronization. However, the  transient time is short compared to the oscillation frequency $\omega = B/N$.  For random initial conditions the oscillation amplitudes even at the boundaries are small. The reason is that the overwhelming majority of states has no overlap with the ground state coherences  such that $C_{01} = \braket{G_{0,0}|\varrho(0)|G_{1,-1}}\ll 1$. However, one may maximize this overlap by choosing $\varrho(0)=\ket{\psi}\bra{\psi}$ as initial state, where $\ket{\psi}$ represents an equal superposition of $\ket{G_{0,0}}$ and $\ket{G_{1,S_z}}$. This suggests that the previously mentioned two-step process may be better suited for actual experimental implementations. Fig.~\ref{fig:AKLT}(c) shows the dynamics for $\ket{\psi} = (\ket{G_{0,0}} + \ket{G_{1,-1}})/\sqrt{2}$ as initial state. As this state is decoherence free, the amplitudes are unaffected by the dissipation and  anti-synchronization is stable. Note, that the transient relaxation time, related to the Liouvillian gap $\Delta$, scales in general exponentially with system size. However, within the ground state subspace $\Delta = 2\gamma$. Thus, in combination with the previously mentioned two-step process, where the ground state manifold can be prepared dissipatively in polyonomial time \cite{zhou2021symmetry} or via a constant-depth quantum circuit \cite{PRXQuantum.4.020315}, the synchronized dynamics can be achieved  efficiently in general.

\emph{Haldane chain.---}The AKLT Hamiltonian~(\ref{eq:Hamiltonian_AKLT}) exhibits spin-1/2 degrees of freedom that are perfectly localized at boundaries and do not interact. In the following we investigate the impact of interactions by decreasing the value of the biquadratic term in Eq.~(\ref{eq:Hamiltonian_AKLT}), i.e., we consider the Hamiltonian
\begin{equation}
H_\varepsilon = H-\varepsilon \sum\limits_{j=1}^{N-1} \left(\vec S_{j}\cdot \vec S_{j+1}\right)^2. 
\end{equation}
For finite values of $\varepsilon$, the Hamiltonian cannot simply be expressed via projection operators and the Lindblad operators $L_{j,j+1}$ induce additional dissipation. Fig.~\ref{fig:Heisenberg}(a) shows the complex eigenvalues of $\mathcal L$ close to the imaginary axis for different values of $\varepsilon$ in the range of $\left[0,1/6\right]$, where $\varepsilon=1/6$ removes the biquadratic term completely, and corresponds to the spin-1 Heisenberg chain (with additional magnetic field). Upon increasing $\varepsilon$ the initially purely imaginary eigenvalues move  away from both the real and the imaginary axis, i.e. the oscillation frequency increases, yet the synchronization is damped. However, the real part remains small and in particular the eigenvalues with second smallest real part also move away from the imaginary axis. Thus, there exist a time range for which all eigenstates but the synchronized ones are damped. Such damped synchronized dynamics   has also been termed metastable synchronization \cite{buca2022algebraic}. 

Stable synchronization may however be restored under certain conditions even for the Heisenberg chain ($\varepsilon=1/6$). To this end we consider the case of $B=0$, such that for $\varepsilon=0$ the ground state is fourfold degenerate. Perturbations to the biquadratic term of the AKLT Hamiltonian partially lift the ground degeneracy such that the $S=0$ state is energetically distinct from the states within the $S=1$ subspace. In the following, we will refer to both the fourfold degenerate ground state for $\varepsilon=0$ as well as the threefold degenerate subspace together with the ground state for $\varepsilon\neq 0$ as the ground state subspace. As $H_\varepsilon$ and the dissipator $L_\mathrm{G}$ preserve the total angular momentum, the dynamics is confined to their respective total angular momentum subspace for $\kappa=0$. Then, there  exists again a dynamical symmetry operator connecting the threefold degenerate subspace ($S=1$) with the $S=0$ state. Similar to the previous discussion, this results in perfect anti-synchronization if the initial state is chosen to be within the  ground state subspace. Figs.~\ref{fig:Heisenberg}(b) and (c) show the time evolution of the  transverse spin $\braket{S_j^\mathrm{x}}(t)$ for a chain of length $N=6$ for $\varepsilon=0.1$ and $\varepsilon=1/6$, respectively. The dynamics show perfect anti-synchronization for the infinite temperature state within the ground state subspace as initial state. The oscillation frequency in Fig.~\ref{fig:Heisenberg} (c) is larger compared to (b) as the energy gap between the $S=1$ and $S=0$ subspaces opens. 

A few remarks are in order: First, the synchronization observed in Figs.~\ref{fig:Heisenberg}(b) and (c) is distinct from regular  coherent dynamics: While without dissipation, coherent (anti-phase) oscillations are still present, there also exists an additional constant shift depending on the initial conditions: different locally acting  observables may not exhibit the same shift, so in the strict definition of Eq.~(\ref{eq:SynchObs}) they are not synchronized. Open system dynamics are thus necessary for perfect (anti-)synchronization even within the ground state subspace of the Haldane chain. Second, perfect anti-synchronization in the Haldane chain ($\varepsilon\neq 0$) is possible without an additional magnetic field ($B=0$), which demonstrates the significance of microscopic details for synchronization even for systems belonging to the same (Haldane) phase. Third, the Haldane phase is protected as long as any \emph{one} of three symmetries is preserved \cite{PhysRevB.81.064439, PhysRevB.85.075125}. That is time-reversal symmetry, link inversion symmetry (lattice inversion about the center of a bond), and $\mathbb Z_2\times \mathbb Z_2$ symmetry ($\pi$ rotations about two orthogonal axes). As a consequence, synchronization within the ground state subspace is robust even if the inversion symmetry is broken, for example by perturbing the interactions between neighboring spins or via an inhomogenous magnetic field \cite{supp}. This is in clear contrast to previous spin chain models \cite{Tindall2020, buca2022algebraic, schmolke2022noise} where permutation symmetry of the Liouvillian is necessary for synchronized dynamics of local observables as defined in Eq.~(\ref{eq:SynchObs}). \rev{In particular, synchronization induced by permutation symmetries becomes, in general, unstable upon symmetry breaking perturbations (i.e., the purely imaginary eigenvalues acquire a negative real part), whereas it remains stable in the presence of symmetry-protected topological order.} However, additional single site-dissipation or decoherence is not protected by the topology \rev{and will render synchronization metastable, similar to other spin chains with dynamical symmetries}. 

\emph{Conclusions.---}We have shown that it is possible to synchronize the \emph{fractionalized} spin degrees of freedom in the spin-1 AKLT chain via engineered dissipation and an external magnetic field. The observed synchronization is stable and topologically protected. While  perturbations to the biquadratic term result in an additional dissipation channel, stable synchronization \rev{is} restored within the ground state subspace of the chain even without magnetic field via a single global spin lowering operator. \rev{Given recent experimental advancements in the preparation of the ground state of the AKLT model on a digital quantum computer \cite{PRXQuantum.4.020315} and the Heisenberg chain in cold atoms \cite{sompet2022realizing}, coupled with the experimental capability to implement collective decay processes on these platforms (either through a combination of ancillary systems and measurements \cite{PRXQuantum.4.010324} or via a collective cavity mode in the `bad cavity' limit \cite{PhysRevX.6.011025, angerer2018superradiant}) the prospect of realizing topological synchronization seems attainable in the near future \cite{supp}.} Our results illuminate on the possibility to utilize dissipation in order to control the dynamics of fractionalized degrees of freedom, not only to prepare them\rev{, and provide a pathway to topologically induced quantum synchronization.}

\emph{Acknowledgements.---}The authors acknowledge useful discussions with Samuel J. Garratt and Ravi K. Naik.  C.~W.~W. was supported by the Deutsche Forschungsgemeinschaft (DFG, German Research Foundation), Project No. 496502542 (WA 5170/1-1), and J.E.M. by the Quantum Materials program under the Director, Office of Science, Office of Basic Energy Sciences, Materials Sciences and Engineering Division, of the U.S. Department of Energy, Contract No. DE-AC02-05CH11231.  Both authors received additional support from a Simons Investigator award.


\begin{thebibliography}{75}%
\makeatletter
\providecommand \@ifxundefined [1]{%
 \@ifx{#1\undefined}
}%
\providecommand \@ifnum [1]{%
 \ifnum #1\expandafter \@firstoftwo
 \else \expandafter \@secondoftwo
 \fi
}%
\providecommand \@ifx [1]{%
 \ifx #1\expandafter \@firstoftwo
 \else \expandafter \@secondoftwo
 \fi
}%
\providecommand \natexlab [1]{#1}%
\providecommand \enquote  [1]{``#1''}%
\providecommand \bibnamefont  [1]{#1}%
\providecommand \bibfnamefont [1]{#1}%
\providecommand \citenamefont [1]{#1}%
\providecommand \href@noop [0]{\@secondoftwo}%
\providecommand \href [0]{\begingroup \@sanitize@url \@href}%
\providecommand \@href[1]{\@@startlink{#1}\@@href}%
\providecommand \@@href[1]{\endgroup#1\@@endlink}%
\providecommand \@sanitize@url [0]{\catcode `\\12\catcode `\$12\catcode
  `\&12\catcode `\#12\catcode `\^12\catcode `\_12\catcode `\%12\relax}%
\providecommand \@@startlink[1]{}%
\providecommand \@@endlink[0]{}%
\providecommand \url  [0]{\begingroup\@sanitize@url \@url }%
\providecommand \@url [1]{\endgroup\@href {#1}{\urlprefix }}%
\providecommand \urlprefix  [0]{URL }%
\providecommand \Eprint [0]{\href }%
\providecommand \doibase [0]{https://doi.org/}%
\providecommand \selectlanguage [0]{\@gobble}%
\providecommand \bibinfo  [0]{\@secondoftwo}%
\providecommand \bibfield  [0]{\@secondoftwo}%
\providecommand \translation [1]{[#1]}%
\providecommand \BibitemOpen [0]{}%
\providecommand \bibitemStop [0]{}%
\providecommand \bibitemNoStop [0]{.\EOS\space}%
\providecommand \EOS [0]{\spacefactor3000\relax}%
\providecommand \BibitemShut  [1]{\csname bibitem#1\endcsname}%
\let\auto@bib@innerbib\@empty
\bibitem [{\citenamefont {Pikovsky}\ \emph {et~al.}(2003)\citenamefont
  {Pikovsky}, \citenamefont {Kurths}, \citenamefont {Rosenblum},\ and\
  \citenamefont {Kurths}}]{pikovsky2003synchronization}%
  \BibitemOpen
  \bibfield  {author} {\bibinfo {author} {\bibfnamefont {A.}~\bibnamefont
  {Pikovsky}}, \bibinfo {author} {\bibfnamefont {J.}~\bibnamefont {Kurths}},
  \bibinfo {author} {\bibfnamefont {M.}~\bibnamefont {Rosenblum}},\ and\
  \bibinfo {author} {\bibfnamefont {J.}~\bibnamefont {Kurths}},\ }\href@noop {}
  {\emph {\bibinfo {title} {Synchronization: a universal concept in nonlinear
  sciences}}}\ (\bibinfo  {publisher} {Cambridge university press},\ \bibinfo
  {year} {Cambridge University Press, 2003})\BibitemShut {NoStop}%
\bibitem [{\citenamefont {Strogatz}(2018)}]{StrogatzBook2018}%
  \BibitemOpen
  \bibfield  {author} {\bibinfo {author} {\bibfnamefont {S.~H.}\ \bibnamefont
  {Strogatz}},\ }\href@noop {} {\emph {\bibinfo {title} {Nonlinear dynamics and
  chaos: with applications to physics, biology, chemistry, and engineering}}}\
  (\bibinfo  {publisher} {CRC Press},\ \bibinfo {year} {2018})\BibitemShut
  {NoStop}%
\bibitem [{\citenamefont {Strogatz}\ and\ \citenamefont
  {Stewart}(1993)}]{Strogatz1993}%
  \BibitemOpen
  \bibfield  {author} {\bibinfo {author} {\bibfnamefont {S.~H.}\ \bibnamefont
  {Strogatz}}\ and\ \bibinfo {author} {\bibfnamefont {I.}~\bibnamefont
  {Stewart}},\ }\bibfield  {title} {\bibinfo {title} {Coupled oscillators and
  biological synchronization},\ }\href@noop {} {\bibfield  {journal} {\bibinfo
  {journal} {Sci. Am.}\ }\textbf {\bibinfo {volume} {269}},\ \bibinfo {pages}
  {102} (\bibinfo {year} {1993})}\BibitemShut {NoStop}%
\bibitem [{\citenamefont {Rosenblum}\ and\ \citenamefont
  {Pikovsky}(2003)}]{Rosenblum2003}%
  \BibitemOpen
  \bibfield  {author} {\bibinfo {author} {\bibfnamefont {M.}~\bibnamefont
  {Rosenblum}}\ and\ \bibinfo {author} {\bibfnamefont {A.}~\bibnamefont
  {Pikovsky}},\ }\bibfield  {title} {\bibinfo {title} {Synchronization: from
  pendulum clocks to chaotic lasers and chemical oscillators},\ }\href
  {https://doi.org/10.1080/00107510310001603129} {\bibfield  {journal}
  {\bibinfo  {journal} {Contemp. Phys.}\ }\textbf {\bibinfo {volume} {44}},\
  \bibinfo {pages} {401} (\bibinfo {year} {2003})}\BibitemShut {NoStop}%
\bibitem [{\citenamefont {Arenas}\ \emph {et~al.}(2008)\citenamefont {Arenas},
  \citenamefont {D{\'\i}az-Guilera}, \citenamefont {Kurths}, \citenamefont
  {Moreno},\ and\ \citenamefont {Zhou}}]{arenas2008synchronization}%
  \BibitemOpen
  \bibfield  {author} {\bibinfo {author} {\bibfnamefont {A.}~\bibnamefont
  {Arenas}}, \bibinfo {author} {\bibfnamefont {A.}~\bibnamefont
  {D{\'\i}az-Guilera}}, \bibinfo {author} {\bibfnamefont {J.}~\bibnamefont
  {Kurths}}, \bibinfo {author} {\bibfnamefont {Y.}~\bibnamefont {Moreno}},\
  and\ \bibinfo {author} {\bibfnamefont {C.}~\bibnamefont {Zhou}},\ }\bibfield
  {title} {\bibinfo {title} {Synchronization in complex networks},\ }\href
  {https://doi.org/https://doi.org/10.1016/j.physrep.2008.09.002} {\bibfield
  {journal} {\bibinfo  {journal} {Phys. Rep.}\ }\textbf {\bibinfo {volume}
  {469}},\ \bibinfo {pages} {93} (\bibinfo {year} {2008})}\BibitemShut
  {NoStop}%
\bibitem [{\citenamefont {Thornburg}\ \emph {et~al.}(1997)\citenamefont
  {Thornburg}, \citenamefont {M{\"o}ller}, \citenamefont {Roy}, \citenamefont
  {Carr}, \citenamefont {Li},\ and\ \citenamefont
  {Erneux}}]{thornburg1997chaos}%
  \BibitemOpen
  \bibfield  {author} {\bibinfo {author} {\bibfnamefont {K.}~\bibnamefont
  {Thornburg}}, \bibinfo {author} {\bibfnamefont {M.}~\bibnamefont
  {M{\"o}ller}}, \bibinfo {author} {\bibfnamefont {R.}~\bibnamefont {Roy}},
  \bibinfo {author} {\bibfnamefont {T.}~\bibnamefont {Carr}}, \bibinfo {author}
  {\bibfnamefont {R.-D.}\ \bibnamefont {Li}},\ and\ \bibinfo {author}
  {\bibfnamefont {T.}~\bibnamefont {Erneux}},\ }\bibfield  {title} {\bibinfo
  {title} {Chaos and coherence in coupled lasers},\ }\href
  {https://link.aps.org/doi/10.1103/PhysRevE.55.3865} {\bibfield  {journal}
  {\bibinfo  {journal} {Phys. Rev. E}\ }\textbf {\bibinfo {volume} {55}},\
  \bibinfo {pages} {3865} (\bibinfo {year} {1997})}\BibitemShut {NoStop}%
\bibitem [{\citenamefont {Lynch}\ and\ \citenamefont
  {York}(1995)}]{lynch1995mode}%
  \BibitemOpen
  \bibfield  {author} {\bibinfo {author} {\bibfnamefont {J.~J.}\ \bibnamefont
  {Lynch}}\ and\ \bibinfo {author} {\bibfnamefont {R.~A.}\ \bibnamefont
  {York}},\ }\bibfield  {title} {\bibinfo {title} {A mode locked array of
  coupled phase locked loops},\ }\href {https://doi.org/10.1109/75.392278}
  {\bibfield  {journal} {\bibinfo  {journal} {IEEE Microw. Guide Wave Lett.}\
  }\textbf {\bibinfo {volume} {5}},\ \bibinfo {pages} {213} (\bibinfo {year}
  {1995})}\BibitemShut {NoStop}%
\bibitem [{\citenamefont {Cawthorne}\ \emph {et~al.}(1999)\citenamefont
  {Cawthorne}, \citenamefont {Barbara}, \citenamefont {Shitov}, \citenamefont
  {Lobb}, \citenamefont {Wiesenfeld},\ and\ \citenamefont
  {Zangwill}}]{cawthorne1999synchronized}%
  \BibitemOpen
  \bibfield  {author} {\bibinfo {author} {\bibfnamefont {A.}~\bibnamefont
  {Cawthorne}}, \bibinfo {author} {\bibfnamefont {P.}~\bibnamefont {Barbara}},
  \bibinfo {author} {\bibfnamefont {S.}~\bibnamefont {Shitov}}, \bibinfo
  {author} {\bibfnamefont {C.}~\bibnamefont {Lobb}}, \bibinfo {author}
  {\bibfnamefont {K.}~\bibnamefont {Wiesenfeld}},\ and\ \bibinfo {author}
  {\bibfnamefont {A.}~\bibnamefont {Zangwill}},\ }\bibfield  {title} {\bibinfo
  {title} {Synchronized oscillations in josephson junction arrays: The role of
  distributed coupling},\ }\href
  {https://link.aps.org/doi/10.1103/PhysRevB.60.7575} {\bibfield  {journal}
  {\bibinfo  {journal} {Phys. Rev. B}\ }\textbf {\bibinfo {volume} {60}},\
  \bibinfo {pages} {7575} (\bibinfo {year} {1999})}\BibitemShut {NoStop}%
\bibitem [{\citenamefont {Fazio}\ and\ \citenamefont {Van
  Der~Zant}(2001)}]{fazio2001quantum}%
  \BibitemOpen
  \bibfield  {author} {\bibinfo {author} {\bibfnamefont {R.}~\bibnamefont
  {Fazio}}\ and\ \bibinfo {author} {\bibfnamefont {H.}~\bibnamefont {Van
  Der~Zant}},\ }\bibfield  {title} {\bibinfo {title} {Quantum phase transitions
  and vortex dynamics in superconducting networks},\ }\href
  {https://doi.org/https://doi.org/10.1016/S0370-1573(01)00022-9} {\bibfield
  {journal} {\bibinfo  {journal} {Phys. Rep.}\ }\textbf {\bibinfo {volume}
  {355}},\ \bibinfo {pages} {235} (\bibinfo {year} {2001})}\BibitemShut
  {NoStop}%
\bibitem [{\citenamefont {Slavin}(2009)}]{slavin2009spin}%
  \BibitemOpen
  \bibfield  {author} {\bibinfo {author} {\bibfnamefont {A.}~\bibnamefont
  {Slavin}},\ }\bibfield  {title} {\bibinfo {title} {Spin-torque oscillators
  get in phase},\ }\href
  {https://doi.org/https://doi.org/10.1038/nnano.2009.213} {\bibfield
  {journal} {\bibinfo  {journal} {Nature Nanotech.}\ }\textbf {\bibinfo
  {volume} {4}},\ \bibinfo {pages} {479} (\bibinfo {year} {2009})}\BibitemShut
  {NoStop}%
\bibitem [{\citenamefont {Nishikawa}\ and\ \citenamefont
  {Motter}(2015)}]{Nishikawa_2015}%
  \BibitemOpen
  \bibfield  {author} {\bibinfo {author} {\bibfnamefont {T.}~\bibnamefont
  {Nishikawa}}\ and\ \bibinfo {author} {\bibfnamefont {A.~E.}\ \bibnamefont
  {Motter}},\ }\bibfield  {title} {\bibinfo {title} {Comparative analysis of
  existing models for power-grid synchronization},\ }\href
  {https://doi.org/10.1088/1367-2630/17/1/015012} {\bibfield  {journal}
  {\bibinfo  {journal} {New J. Phys.}\ }\textbf {\bibinfo {volume} {17}},\
  \bibinfo {pages} {015012} (\bibinfo {year} {2015})}\BibitemShut {NoStop}%
\bibitem [{\citenamefont {Bellamy}(1995)}]{bellamy1995digital}%
  \BibitemOpen
  \bibfield  {author} {\bibinfo {author} {\bibfnamefont {J.~C.}\ \bibnamefont
  {Bellamy}},\ }\bibfield  {title} {\bibinfo {title} {Digital network
  synchronization},\ }\href {https://doi.org/10.1109/35.372197} {\bibfield
  {journal} {\bibinfo  {journal} {IEEE Commun. Mag.}\ }\textbf {\bibinfo
  {volume} {33}},\ \bibinfo {pages} {70} (\bibinfo {year} {1995})}\BibitemShut
  {NoStop}%
\bibitem [{\citenamefont {Narula}\ and\ \citenamefont
  {Humphreys}(2018)}]{narula2018requirements}%
  \BibitemOpen
  \bibfield  {author} {\bibinfo {author} {\bibfnamefont {L.}~\bibnamefont
  {Narula}}\ and\ \bibinfo {author} {\bibfnamefont {T.~E.}\ \bibnamefont
  {Humphreys}},\ }\bibfield  {title} {\bibinfo {title} {Requirements for secure
  clock synchronization},\ }\href {https://doi.org/10.1109/JSTSP.2018.2835772}
  {\bibfield  {journal} {\bibinfo  {journal} {IEEE J. Sel. Top. Signal
  Process.}\ }\textbf {\bibinfo {volume} {12}},\ \bibinfo {pages} {749}
  (\bibinfo {year} {2018})}\BibitemShut {NoStop}%
\bibitem [{\citenamefont {Lee}\ and\ \citenamefont
  {Sadeghpour}(2013)}]{Lee2013}%
  \BibitemOpen
  \bibfield  {author} {\bibinfo {author} {\bibfnamefont {T.~E.}\ \bibnamefont
  {Lee}}\ and\ \bibinfo {author} {\bibfnamefont {H.~R.}\ \bibnamefont
  {Sadeghpour}},\ }\bibfield  {title} {\bibinfo {title} {Quantum
  synchronization of quantum van der pol oscillators with trapped ions},\
  }\href {https://doi.org/10.1103/PhysRevLett.111.234101} {\bibfield  {journal}
  {\bibinfo  {journal} {Phys. Rev. Lett.}\ }\textbf {\bibinfo {volume} {111}},\
  \bibinfo {pages} {234101} (\bibinfo {year} {2013})}\BibitemShut {NoStop}%
\bibitem [{\citenamefont {Walter}\ \emph {et~al.}(2014)\citenamefont {Walter},
  \citenamefont {Nunnenkamp},\ and\ \citenamefont
  {Bruder}}]{PhysRevLett.112.094102}%
  \BibitemOpen
  \bibfield  {author} {\bibinfo {author} {\bibfnamefont {S.}~\bibnamefont
  {Walter}}, \bibinfo {author} {\bibfnamefont {A.}~\bibnamefont {Nunnenkamp}},\
  and\ \bibinfo {author} {\bibfnamefont {C.}~\bibnamefont {Bruder}},\
  }\bibfield  {title} {\bibinfo {title} {Quantum synchronization of a driven
  self-sustained oscillator},\ }\href
  {https://doi.org/10.1103/PhysRevLett.112.094102} {\bibfield  {journal}
  {\bibinfo  {journal} {Phys. Rev. Lett.}\ }\textbf {\bibinfo {volume} {112}},\
  \bibinfo {pages} {094102} (\bibinfo {year} {2014})}\BibitemShut {NoStop}%
\bibitem [{\citenamefont {Dutta}\ and\ \citenamefont
  {Cooper}(2019)}]{dutta2019critical}%
  \BibitemOpen
  \bibfield  {author} {\bibinfo {author} {\bibfnamefont {S.}~\bibnamefont
  {Dutta}}\ and\ \bibinfo {author} {\bibfnamefont {N.~R.}\ \bibnamefont
  {Cooper}},\ }\bibfield  {title} {\bibinfo {title} {Critical response of a
  quantum van der pol oscillator},\ }\href
  {https://doi.org/10.1103/PhysRevLett.123.250401} {\bibfield  {journal}
  {\bibinfo  {journal} {Phys. Rev. Lett.}\ }\textbf {\bibinfo {volume} {123}},\
  \bibinfo {pages} {250401} (\bibinfo {year} {2019})}\BibitemShut {NoStop}%
\bibitem [{\citenamefont {Walter}\ \emph {et~al.}(2015)\citenamefont {Walter},
  \citenamefont {Nunnenkamp},\ and\ \citenamefont {Bruder}}]{Walter2015}%
  \BibitemOpen
  \bibfield  {author} {\bibinfo {author} {\bibfnamefont {S.}~\bibnamefont
  {Walter}}, \bibinfo {author} {\bibfnamefont {A.}~\bibnamefont {Nunnenkamp}},\
  and\ \bibinfo {author} {\bibfnamefont {C.}~\bibnamefont {Bruder}},\
  }\bibfield  {title} {\bibinfo {title} {Quantum synchronization of two van der
  pol oscillators},\ }\href {https://doi.org/10.1002/andp.201400144} {\bibfield
   {journal} {\bibinfo  {journal} {Ann. Phys.}\ }\textbf {\bibinfo {volume}
  {527}},\ \bibinfo {pages} {131} (\bibinfo {year} {2015})}\BibitemShut
  {NoStop}%
\bibitem [{\citenamefont {Bastidas}\ \emph {et~al.}(2015)\citenamefont
  {Bastidas}, \citenamefont {Omelchenko}, \citenamefont {Zakharova},
  \citenamefont {Sch{\"o}ll},\ and\ \citenamefont {Brandes}}]{bastidas2015}%
  \BibitemOpen
  \bibfield  {author} {\bibinfo {author} {\bibfnamefont {V.~M.}\ \bibnamefont
  {Bastidas}}, \bibinfo {author} {\bibfnamefont {I.}~\bibnamefont
  {Omelchenko}}, \bibinfo {author} {\bibfnamefont {A.}~\bibnamefont
  {Zakharova}}, \bibinfo {author} {\bibfnamefont {E.}~\bibnamefont
  {Sch{\"o}ll}},\ and\ \bibinfo {author} {\bibfnamefont {T.}~\bibnamefont
  {Brandes}},\ }\bibfield  {title} {\bibinfo {title} {Quantum signatures of
  chimera states},\ }\href
  {https://link.aps.org/doi/10.1103/PhysRevE.92.062924} {\bibfield  {journal}
  {\bibinfo  {journal} {Phys. Rev. E}\ }\textbf {\bibinfo {volume} {92}},\
  \bibinfo {pages} {062924} (\bibinfo {year} {2015})}\BibitemShut {NoStop}%
\bibitem [{\citenamefont {Ishibashi}\ and\ \citenamefont
  {Kanamoto}(2017)}]{PhysRevE.96.052210}%
  \BibitemOpen
  \bibfield  {author} {\bibinfo {author} {\bibfnamefont {K.}~\bibnamefont
  {Ishibashi}}\ and\ \bibinfo {author} {\bibfnamefont {R.}~\bibnamefont
  {Kanamoto}},\ }\bibfield  {title} {\bibinfo {title} {Oscillation collapse in
  coupled quantum van der pol oscillators},\ }\href
  {https://doi.org/10.1103/PhysRevE.96.052210} {\bibfield  {journal} {\bibinfo
  {journal} {Phys. Rev. E}\ }\textbf {\bibinfo {volume} {96}},\ \bibinfo
  {pages} {052210} (\bibinfo {year} {2017})}\BibitemShut {NoStop}%
\bibitem [{\citenamefont {Mari}\ \emph {et~al.}(2013)\citenamefont {Mari},
  \citenamefont {Farace}, \citenamefont {Didier}, \citenamefont {Giovannetti},\
  and\ \citenamefont {Fazio}}]{2013-Fazio-PRL}%
  \BibitemOpen
  \bibfield  {author} {\bibinfo {author} {\bibfnamefont {A.}~\bibnamefont
  {Mari}}, \bibinfo {author} {\bibfnamefont {A.}~\bibnamefont {Farace}},
  \bibinfo {author} {\bibfnamefont {N.}~\bibnamefont {Didier}}, \bibinfo
  {author} {\bibfnamefont {V.}~\bibnamefont {Giovannetti}},\ and\ \bibinfo
  {author} {\bibfnamefont {R.}~\bibnamefont {Fazio}},\ }\bibfield  {title}
  {\bibinfo {title} {{Measures of Quantum Synchronization in Continuous
  Variable Systems}},\ }\href {https://doi.org/10.1103/PhysRevLett.111.103605}
  {\bibfield  {journal} {\bibinfo  {journal} {Phys. Rev. Lett.}\ }\textbf
  {\bibinfo {volume} {111}},\ \bibinfo {pages} {103605} (\bibinfo {year}
  {2013})}\BibitemShut {NoStop}%
\bibitem [{\citenamefont {Thomas}\ and\ \citenamefont
  {Senthilvelan}(2022)}]{PhysRevA.106.012422}%
  \BibitemOpen
  \bibfield  {author} {\bibinfo {author} {\bibfnamefont {N.}~\bibnamefont
  {Thomas}}\ and\ \bibinfo {author} {\bibfnamefont {M.}~\bibnamefont
  {Senthilvelan}},\ }\bibfield  {title} {\bibinfo {title} {Quantum
  synchronization in quadratically coupled quantum van der pol oscillators},\
  }\href {https://doi.org/10.1103/PhysRevA.106.012422} {\bibfield  {journal}
  {\bibinfo  {journal} {Phys. Rev. A}\ }\textbf {\bibinfo {volume} {106}},\
  \bibinfo {pages} {012422} (\bibinfo {year} {2022})}\BibitemShut {NoStop}%
\bibitem [{\citenamefont {Ameri}\ \emph {et~al.}(2015)\citenamefont {Ameri},
  \citenamefont {Eghbali-Arani}, \citenamefont {Mari}, \citenamefont {Farace},
  \citenamefont {Kheirandish}, \citenamefont {Giovannetti},\ and\ \citenamefont
  {Fazio}}]{ameri2015mutual}%
  \BibitemOpen
  \bibfield  {author} {\bibinfo {author} {\bibfnamefont {V.}~\bibnamefont
  {Ameri}}, \bibinfo {author} {\bibfnamefont {M.}~\bibnamefont
  {Eghbali-Arani}}, \bibinfo {author} {\bibfnamefont {A.}~\bibnamefont {Mari}},
  \bibinfo {author} {\bibfnamefont {A.}~\bibnamefont {Farace}}, \bibinfo
  {author} {\bibfnamefont {F.}~\bibnamefont {Kheirandish}}, \bibinfo {author}
  {\bibfnamefont {V.}~\bibnamefont {Giovannetti}},\ and\ \bibinfo {author}
  {\bibfnamefont {R.}~\bibnamefont {Fazio}},\ }\bibfield  {title} {\bibinfo
  {title} {Mutual information as an order parameter for quantum
  synchronization},\ }\href
  {https://link.aps.org/doi/10.1103/PhysRevA.91.012301} {\bibfield  {journal}
  {\bibinfo  {journal} {Phys. Rev. A}\ }\textbf {\bibinfo {volume} {91}},\
  \bibinfo {pages} {012301} (\bibinfo {year} {2015})}\BibitemShut {NoStop}%
\bibitem [{\citenamefont {Sonar}\ \emph {et~al.}(2018)\citenamefont {Sonar},
  \citenamefont {Hajdu{\v{s}}ek}, \citenamefont {Mukherjee}, \citenamefont
  {Fazio}, \citenamefont {Vedral}, \citenamefont {Vinjanampathy},\ and\
  \citenamefont {Kwek}}]{sonar2018squeezing}%
  \BibitemOpen
  \bibfield  {author} {\bibinfo {author} {\bibfnamefont {S.}~\bibnamefont
  {Sonar}}, \bibinfo {author} {\bibfnamefont {M.}~\bibnamefont
  {Hajdu{\v{s}}ek}}, \bibinfo {author} {\bibfnamefont {M.}~\bibnamefont
  {Mukherjee}}, \bibinfo {author} {\bibfnamefont {R.}~\bibnamefont {Fazio}},
  \bibinfo {author} {\bibfnamefont {V.}~\bibnamefont {Vedral}}, \bibinfo
  {author} {\bibfnamefont {S.}~\bibnamefont {Vinjanampathy}},\ and\ \bibinfo
  {author} {\bibfnamefont {L.-C.}\ \bibnamefont {Kwek}},\ }\bibfield  {title}
  {\bibinfo {title} {Squeezing enhances quantum synchronization},\ }\href
  {https://link.aps.org/doi/10.1103/PhysRevLett.120.163601} {\bibfield
  {journal} {\bibinfo  {journal} {Phys. Rev. Lett.}\ }\textbf {\bibinfo
  {volume} {120}},\ \bibinfo {pages} {163601} (\bibinfo {year}
  {2018})}\BibitemShut {NoStop}%
\bibitem [{\citenamefont {Es'~haqi Sani}\ \emph {et~al.}(2020)\citenamefont
  {Es'~haqi Sani}, \citenamefont {Manzano}, \citenamefont {Zambrini},\ and\
  \citenamefont {Fazio}}]{es2020synchronization}%
  \BibitemOpen
  \bibfield  {author} {\bibinfo {author} {\bibfnamefont {N.}~\bibnamefont
  {Es'~haqi Sani}}, \bibinfo {author} {\bibfnamefont {G.}~\bibnamefont
  {Manzano}}, \bibinfo {author} {\bibfnamefont {R.}~\bibnamefont {Zambrini}},\
  and\ \bibinfo {author} {\bibfnamefont {R.}~\bibnamefont {Fazio}},\ }\bibfield
   {title} {\bibinfo {title} {Synchronization along quantum trajectories},\
  }\href {https://link.aps.org/doi/10.1103/PhysRevResearch.2.023101} {\bibfield
   {journal} {\bibinfo  {journal} {Phys. Rev. Research}\ }\textbf {\bibinfo
  {volume} {2}},\ \bibinfo {pages} {023101} (\bibinfo {year}
  {2020})}\BibitemShut {NoStop}%
\bibitem [{\citenamefont {L{\"o}rch}\ \emph {et~al.}(2016)\citenamefont
  {L{\"o}rch}, \citenamefont {Amitai}, \citenamefont {Nunnenkamp},\ and\
  \citenamefont {Bruder}}]{lorch2016genuine}%
  \BibitemOpen
  \bibfield  {author} {\bibinfo {author} {\bibfnamefont {N.}~\bibnamefont
  {L{\"o}rch}}, \bibinfo {author} {\bibfnamefont {E.}~\bibnamefont {Amitai}},
  \bibinfo {author} {\bibfnamefont {A.}~\bibnamefont {Nunnenkamp}},\ and\
  \bibinfo {author} {\bibfnamefont {C.}~\bibnamefont {Bruder}},\ }\bibfield
  {title} {\bibinfo {title} {Genuine quantum signatures in synchronization of
  anharmonic self-oscillators},\ }\href
  {https://link.aps.org/doi/10.1103/PhysRevLett.117.073601} {\bibfield
  {journal} {\bibinfo  {journal} {Phys. Rev. Lett.}\ }\textbf {\bibinfo
  {volume} {117}},\ \bibinfo {pages} {073601} (\bibinfo {year}
  {2016})}\BibitemShut {NoStop}%
\bibitem [{\citenamefont {Davis-Tilley}\ \emph {et~al.}(2018)\citenamefont
  {Davis-Tilley}, \citenamefont {Teoh},\ and\ \citenamefont
  {Armour}}]{Davis2018}%
  \BibitemOpen
  \bibfield  {author} {\bibinfo {author} {\bibfnamefont {C.}~\bibnamefont
  {Davis-Tilley}}, \bibinfo {author} {\bibfnamefont {C.~K.}\ \bibnamefont
  {Teoh}},\ and\ \bibinfo {author} {\bibfnamefont {A.~D.}\ \bibnamefont
  {Armour}},\ }\bibfield  {title} {\bibinfo {title} {Dynamics of many-body
  quantum synchronisation},\ }\href {https://doi.org/10.1088/1367-2630/aae947}
  {\bibfield  {journal} {\bibinfo  {journal} {New J. Phys.}\ }\textbf {\bibinfo
  {volume} {20}},\ \bibinfo {pages} {113002} (\bibinfo {year}
  {2018})}\BibitemShut {NoStop}%
\bibitem [{\citenamefont {Amitai}\ \emph {et~al.}(2018)\citenamefont {Amitai},
  \citenamefont {Koppenh\"ofer}, \citenamefont {L\"orch},\ and\ \citenamefont
  {Bruder}}]{amitai2018quantum}%
  \BibitemOpen
  \bibfield  {author} {\bibinfo {author} {\bibfnamefont {E.}~\bibnamefont
  {Amitai}}, \bibinfo {author} {\bibfnamefont {M.}~\bibnamefont
  {Koppenh\"ofer}}, \bibinfo {author} {\bibfnamefont {N.}~\bibnamefont
  {L\"orch}},\ and\ \bibinfo {author} {\bibfnamefont {C.}~\bibnamefont
  {Bruder}},\ }\bibfield  {title} {\bibinfo {title} {Quantum effects in
  amplitude death of coupled anharmonic self-oscillators},\ }\href
  {https://doi.org/10.1103/PhysRevE.97.052203} {\bibfield  {journal} {\bibinfo
  {journal} {Phys. Rev. E}\ }\textbf {\bibinfo {volume} {97}},\ \bibinfo
  {pages} {052203} (\bibinfo {year} {2018})}\BibitemShut {NoStop}%
\bibitem [{\citenamefont {Mok}\ \emph {et~al.}(2020)\citenamefont {Mok},
  \citenamefont {Kwek},\ and\ \citenamefont
  {Heimonen}}]{mok2020synchronization}%
  \BibitemOpen
  \bibfield  {author} {\bibinfo {author} {\bibfnamefont {W.-K.}\ \bibnamefont
  {Mok}}, \bibinfo {author} {\bibfnamefont {L.-C.}\ \bibnamefont {Kwek}},\ and\
  \bibinfo {author} {\bibfnamefont {H.}~\bibnamefont {Heimonen}},\ }\bibfield
  {title} {\bibinfo {title} {Synchronization boost with single-photon
  dissipation in the deep quantum regime},\ }\href
  {https://doi.org/10.1103/PhysRevResearch.2.033422} {\bibfield  {journal}
  {\bibinfo  {journal} {Phys. Rev. Res.}\ }\textbf {\bibinfo {volume} {2}},\
  \bibinfo {pages} {033422} (\bibinfo {year} {2020})}\BibitemShut {NoStop}%
\bibitem [{\citenamefont {W\"achtler}\ and\ \citenamefont
  {Platero}(2023)}]{PhysRevResearch.5.023021}%
  \BibitemOpen
  \bibfield  {author} {\bibinfo {author} {\bibfnamefont {C.~W.}\ \bibnamefont
  {W\"achtler}}\ and\ \bibinfo {author} {\bibfnamefont {G.}~\bibnamefont
  {Platero}},\ }\bibfield  {title} {\bibinfo {title} {Topological
  synchronization of quantum van der pol oscillators},\ }\href
  {https://doi.org/10.1103/PhysRevResearch.5.023021} {\bibfield  {journal}
  {\bibinfo  {journal} {Phys. Rev. Res.}\ }\textbf {\bibinfo {volume} {5}},\
  \bibinfo {pages} {023021} (\bibinfo {year} {2023})}\BibitemShut {NoStop}%
\bibitem [{\citenamefont {Delmonte}\ \emph {et~al.}(2023)\citenamefont
  {Delmonte}, \citenamefont {Romito}, \citenamefont {Santoro},\ and\
  \citenamefont {Fazio}}]{delmonte2023quantum}%
  \BibitemOpen
  \bibfield  {author} {\bibinfo {author} {\bibfnamefont {A.}~\bibnamefont
  {Delmonte}}, \bibinfo {author} {\bibfnamefont {A.}~\bibnamefont {Romito}},
  \bibinfo {author} {\bibfnamefont {G.~E.}\ \bibnamefont {Santoro}},\ and\
  \bibinfo {author} {\bibfnamefont {R.}~\bibnamefont {Fazio}},\ }\bibfield
  {title} {\bibinfo {title} {Quantum effects on the synchronization dynamics of
  the kuramoto model},\ }\href@noop {} {\bibfield  {journal} {\bibinfo
  {journal} {arXiv preprint arXiv:2306.09956}\ } (\bibinfo {year}
  {2023})}\BibitemShut {NoStop}%
\bibitem [{\citenamefont {Shen}\ \emph {et~al.}(2023)\citenamefont {Shen},
  \citenamefont {Soh}, \citenamefont {Fan},\ and\ \citenamefont
  {Kwek}}]{shen2023enhancing}%
  \BibitemOpen
  \bibfield  {author} {\bibinfo {author} {\bibfnamefont {Y.}~\bibnamefont
  {Shen}}, \bibinfo {author} {\bibfnamefont {H.~Y.}\ \bibnamefont {Soh}},
  \bibinfo {author} {\bibfnamefont {W.}~\bibnamefont {Fan}},\ and\ \bibinfo
  {author} {\bibfnamefont {L.-C.}\ \bibnamefont {Kwek}},\ }\bibfield  {title}
  {\bibinfo {title} {Enhancing quantum synchronization through homodyne
  measurement and squeezing},\ }\href@noop {} {\bibfield  {journal} {\bibinfo
  {journal} {arXiv preprint arXiv:2302.13465}\ } (\bibinfo {year}
  {2023})}\BibitemShut {NoStop}%
\bibitem [{\citenamefont {Roulet}\ and\ \citenamefont
  {Bruder}(2018{\natexlab{a}})}]{PhysRevLett.121.053601}%
  \BibitemOpen
  \bibfield  {author} {\bibinfo {author} {\bibfnamefont {A.}~\bibnamefont
  {Roulet}}\ and\ \bibinfo {author} {\bibfnamefont {C.}~\bibnamefont
  {Bruder}},\ }\bibfield  {title} {\bibinfo {title} {Synchronizing the smallest
  possible system},\ }\href {https://doi.org/10.1103/PhysRevLett.121.053601}
  {\bibfield  {journal} {\bibinfo  {journal} {Phys. Rev. Lett.}\ }\textbf
  {\bibinfo {volume} {121}},\ \bibinfo {pages} {053601} (\bibinfo {year}
  {2018}{\natexlab{a}})}\BibitemShut {NoStop}%
\bibitem [{\citenamefont {Roulet}\ and\ \citenamefont
  {Bruder}(2018{\natexlab{b}})}]{PhysRevLett.121.063601}%
  \BibitemOpen
  \bibfield  {author} {\bibinfo {author} {\bibfnamefont {A.}~\bibnamefont
  {Roulet}}\ and\ \bibinfo {author} {\bibfnamefont {C.}~\bibnamefont
  {Bruder}},\ }\bibfield  {title} {\bibinfo {title} {Quantum synchronization
  and entanglement generation},\ }\href
  {https://link.aps.org/doi/10.1103/PhysRevLett.121.063601} {\bibfield
  {journal} {\bibinfo  {journal} {Phys. Rev. Lett.}\ }\textbf {\bibinfo
  {volume} {121}},\ \bibinfo {pages} {063601} (\bibinfo {year}
  {2018}{\natexlab{b}})}\BibitemShut {NoStop}%
\bibitem [{\citenamefont {Daniel}\ \emph {et~al.}(2023)\citenamefont {Daniel},
  \citenamefont {Bruder},\ and\ \citenamefont
  {Koppenh\"ofer}}]{daniel2023geometric}%
  \BibitemOpen
  \bibfield  {author} {\bibinfo {author} {\bibfnamefont {A.}~\bibnamefont
  {Daniel}}, \bibinfo {author} {\bibfnamefont {C.}~\bibnamefont {Bruder}},\
  and\ \bibinfo {author} {\bibfnamefont {M.}~\bibnamefont {Koppenh\"ofer}},\
  }\bibfield  {title} {\bibinfo {title} {Geometric phase in quantum
  synchronization},\ }\href {https://doi.org/10.1103/PhysRevResearch.5.023182}
  {\bibfield  {journal} {\bibinfo  {journal} {Phys. Rev. Res.}\ }\textbf
  {\bibinfo {volume} {5}},\ \bibinfo {pages} {023182} (\bibinfo {year}
  {2023})}\BibitemShut {NoStop}%
\bibitem [{\citenamefont {Schmolke}\ and\ \citenamefont
  {Lutz}(2022)}]{schmolke2022noise}%
  \BibitemOpen
  \bibfield  {author} {\bibinfo {author} {\bibfnamefont {F.}~\bibnamefont
  {Schmolke}}\ and\ \bibinfo {author} {\bibfnamefont {E.}~\bibnamefont
  {Lutz}},\ }\bibfield  {title} {\bibinfo {title} {Noise-induced quantum
  synchronization},\ }\href {https://doi.org/10.1103/PhysRevLett.129.250601}
  {\bibfield  {journal} {\bibinfo  {journal} {Phys. Rev. Lett.}\ }\textbf
  {\bibinfo {volume} {129}},\ \bibinfo {pages} {250601} (\bibinfo {year}
  {2022})}\BibitemShut {NoStop}%
\bibitem [{\citenamefont {Schmolke}\ and\ \citenamefont
  {Lutz}(2023)}]{schmolke2023measurement}%
  \BibitemOpen
  \bibfield  {author} {\bibinfo {author} {\bibfnamefont {F.}~\bibnamefont
  {Schmolke}}\ and\ \bibinfo {author} {\bibfnamefont {E.}~\bibnamefont
  {Lutz}},\ }\bibfield  {title} {\bibinfo {title} {Measurement-induced quantum
  synchronization and multiplexing},\ }\href@noop {} {\bibfield  {journal}
  {\bibinfo  {journal} {arXiv preprint arXiv:2306.12986}\ } (\bibinfo {year}
  {2023})}\BibitemShut {NoStop}%
\bibitem [{\citenamefont {L\"orch}\ \emph {et~al.}(2017)\citenamefont
  {L\"orch}, \citenamefont {Nigg}, \citenamefont {Nunnenkamp}, \citenamefont
  {Tiwari},\ and\ \citenamefont {Bruder}}]{PhysRevLett.118.243602}%
  \BibitemOpen
  \bibfield  {author} {\bibinfo {author} {\bibfnamefont {N.}~\bibnamefont
  {L\"orch}}, \bibinfo {author} {\bibfnamefont {S.~E.}\ \bibnamefont {Nigg}},
  \bibinfo {author} {\bibfnamefont {A.}~\bibnamefont {Nunnenkamp}}, \bibinfo
  {author} {\bibfnamefont {R.~P.}\ \bibnamefont {Tiwari}},\ and\ \bibinfo
  {author} {\bibfnamefont {C.}~\bibnamefont {Bruder}},\ }\bibfield  {title}
  {\bibinfo {title} {Quantum synchronization blockade: Energy quantization
  hinders synchronization of identical oscillators},\ }\href
  {https://doi.org/10.1103/PhysRevLett.118.243602} {\bibfield  {journal}
  {\bibinfo  {journal} {Phys. Rev. Lett.}\ }\textbf {\bibinfo {volume} {118}},\
  \bibinfo {pages} {243602} (\bibinfo {year} {2017})}\BibitemShut {NoStop}%
\bibitem [{\citenamefont {Tan}\ \emph {et~al.}(2022)\citenamefont {Tan},
  \citenamefont {Bruder},\ and\ \citenamefont {Koppenh{\"o}fer}}]{tan2022half}%
  \BibitemOpen
  \bibfield  {author} {\bibinfo {author} {\bibfnamefont {R.}~\bibnamefont
  {Tan}}, \bibinfo {author} {\bibfnamefont {C.}~\bibnamefont {Bruder}},\ and\
  \bibinfo {author} {\bibfnamefont {M.}~\bibnamefont {Koppenh{\"o}fer}},\
  }\bibfield  {title} {\bibinfo {title} {Half-integer vs. integer effects in
  quantum synchronization of spin systems},\ }\href
  {https://quantum-journal.org/papers/q-2022-12-29-885/} {\bibfield  {journal}
  {\bibinfo  {journal} {Quantum}\ }\textbf {\bibinfo {volume} {6}},\ \bibinfo
  {pages} {885} (\bibinfo {year} {2022})}\BibitemShut {NoStop}%
\bibitem [{\citenamefont {Koppenh\"ofer}\ \emph {et~al.}(2020)\citenamefont
  {Koppenh\"ofer}, \citenamefont {Bruder},\ and\ \citenamefont
  {Roulet}}]{PhysRevResearch.2.023026}%
  \BibitemOpen
  \bibfield  {author} {\bibinfo {author} {\bibfnamefont {M.}~\bibnamefont
  {Koppenh\"ofer}}, \bibinfo {author} {\bibfnamefont {C.}~\bibnamefont
  {Bruder}},\ and\ \bibinfo {author} {\bibfnamefont {A.}~\bibnamefont
  {Roulet}},\ }\bibfield  {title} {\bibinfo {title} {Quantum synchronization on
  the ibm q system},\ }\href {https://doi.org/10.1103/PhysRevResearch.2.023026}
  {\bibfield  {journal} {\bibinfo  {journal} {Phys. Rev. Research}\ }\textbf
  {\bibinfo {volume} {2}},\ \bibinfo {pages} {023026} (\bibinfo {year}
  {2020})}\BibitemShut {NoStop}%
\bibitem [{\citenamefont {Pljonkin}(2021)}]{pljonkin2021vulnerability}%
  \BibitemOpen
  \bibfield  {author} {\bibinfo {author} {\bibfnamefont {A.}~\bibnamefont
  {Pljonkin}},\ }\bibfield  {title} {\bibinfo {title} {Vulnerability of the
  synchronization process in the quantum key distribution system},\ }in\
  \href@noop {} {\emph {\bibinfo {booktitle} {Research Anthology on
  Advancements in Quantum Technology}}}\ (\bibinfo  {publisher} {IGI Global},\
  \bibinfo {year} {2021})\ pp.\ \bibinfo {pages} {345--354}\BibitemShut
  {NoStop}%
\bibitem [{\citenamefont {Pljonkin}\ \emph {et~al.}(2017)\citenamefont
  {Pljonkin}, \citenamefont {Rumyantsev},\ and\ \citenamefont
  {Singh}}]{pljonkin2017synchronization}%
  \BibitemOpen
  \bibfield  {author} {\bibinfo {author} {\bibfnamefont {A.}~\bibnamefont
  {Pljonkin}}, \bibinfo {author} {\bibfnamefont {K.}~\bibnamefont
  {Rumyantsev}},\ and\ \bibinfo {author} {\bibfnamefont {P.}~\bibnamefont
  {Singh}},\ }\bibfield  {title} {\bibinfo {title} {Synchronization in quantum
  key distribution systems},\ }\href
  {https://doi.org/10.3390/cryptography1030018} {\bibfield  {journal} {\bibinfo
   {journal} {Cryptography}\ }\textbf {\bibinfo {volume} {1}},\ \bibinfo
  {pages} {18} (\bibinfo {year} {2017})}\BibitemShut {NoStop}%
\bibitem [{\citenamefont {Liu}\ and\ \citenamefont
  {Yin}(2019)}]{liu2019secure}%
  \BibitemOpen
  \bibfield  {author} {\bibinfo {author} {\bibfnamefont {P.}~\bibnamefont
  {Liu}}\ and\ \bibinfo {author} {\bibfnamefont {H.-L.}\ \bibnamefont {Yin}},\
  }\bibfield  {title} {\bibinfo {title} {Secure and efficient synchronization
  scheme for quantum key distribution},\ }\href
  {https://doi.org/10.1364/OSAC.2.002883} {\bibfield  {journal} {\bibinfo
  {journal} {OSA Continuum}\ }\textbf {\bibinfo {volume} {2}},\ \bibinfo
  {pages} {2883} (\bibinfo {year} {2019})}\BibitemShut {NoStop}%
\bibitem [{\citenamefont {Agnesi}\ \emph {et~al.}(2020)\citenamefont {Agnesi},
  \citenamefont {Avesani}, \citenamefont {Calderaro}, \citenamefont {Stanco},
  \citenamefont {Foletto}, \citenamefont {Zahidy}, \citenamefont {Scriminich},
  \citenamefont {Vedovato}, \citenamefont {Vallone},\ and\ \citenamefont
  {Villoresi}}]{agnesi2020simple}%
  \BibitemOpen
  \bibfield  {author} {\bibinfo {author} {\bibfnamefont {C.}~\bibnamefont
  {Agnesi}}, \bibinfo {author} {\bibfnamefont {M.}~\bibnamefont {Avesani}},
  \bibinfo {author} {\bibfnamefont {L.}~\bibnamefont {Calderaro}}, \bibinfo
  {author} {\bibfnamefont {A.}~\bibnamefont {Stanco}}, \bibinfo {author}
  {\bibfnamefont {G.}~\bibnamefont {Foletto}}, \bibinfo {author} {\bibfnamefont
  {M.}~\bibnamefont {Zahidy}}, \bibinfo {author} {\bibfnamefont
  {A.}~\bibnamefont {Scriminich}}, \bibinfo {author} {\bibfnamefont
  {F.}~\bibnamefont {Vedovato}}, \bibinfo {author} {\bibfnamefont
  {G.}~\bibnamefont {Vallone}},\ and\ \bibinfo {author} {\bibfnamefont
  {P.}~\bibnamefont {Villoresi}},\ }\bibfield  {title} {\bibinfo {title}
  {Simple quantum key distribution with qubit-based synchronization and a
  self-compensating polarization encoder},\ }\href
  {https://doi.org/10.1364/OPTICA.381013} {\bibfield  {journal} {\bibinfo
  {journal} {Optica}\ }\textbf {\bibinfo {volume} {7}},\ \bibinfo {pages} {284}
  (\bibinfo {year} {2020})}\BibitemShut {NoStop}%
\bibitem [{\citenamefont {Jaseem}\ \emph {et~al.}(2020)\citenamefont {Jaseem},
  \citenamefont {Hajdu\ifmmode~\check{s}\else \v{s}\fi{}ek}, \citenamefont
  {Vedral}, \citenamefont {Fazio}, \citenamefont {Kwek},\ and\ \citenamefont
  {Vinjanampathy}}]{jaseem2020quantum}%
  \BibitemOpen
  \bibfield  {author} {\bibinfo {author} {\bibfnamefont {N.}~\bibnamefont
  {Jaseem}}, \bibinfo {author} {\bibfnamefont {M.}~\bibnamefont
  {Hajdu\ifmmode~\check{s}\else \v{s}\fi{}ek}}, \bibinfo {author}
  {\bibfnamefont {V.}~\bibnamefont {Vedral}}, \bibinfo {author} {\bibfnamefont
  {R.}~\bibnamefont {Fazio}}, \bibinfo {author} {\bibfnamefont {L.-C.}\
  \bibnamefont {Kwek}},\ and\ \bibinfo {author} {\bibfnamefont
  {S.}~\bibnamefont {Vinjanampathy}},\ }\bibfield  {title} {\bibinfo {title}
  {Quantum synchronization in nanoscale heat engines},\ }\href
  {https://doi.org/10.1103/PhysRevE.101.020201} {\bibfield  {journal} {\bibinfo
   {journal} {Phys. Rev. E}\ }\textbf {\bibinfo {volume} {101}},\ \bibinfo
  {pages} {020201} (\bibinfo {year} {2020})}\BibitemShut {NoStop}%
\bibitem [{\citenamefont {Murtadho}\ \emph
  {et~al.}(2023{\natexlab{a}})\citenamefont {Murtadho}, \citenamefont
  {Vinjanampathy},\ and\ \citenamefont {Thingna}}]{murtadho2023cooperation}%
  \BibitemOpen
  \bibfield  {author} {\bibinfo {author} {\bibfnamefont {T.}~\bibnamefont
  {Murtadho}}, \bibinfo {author} {\bibfnamefont {S.}~\bibnamefont
  {Vinjanampathy}},\ and\ \bibinfo {author} {\bibfnamefont {J.}~\bibnamefont
  {Thingna}},\ }\bibfield  {title} {\bibinfo {title} {Cooperation and
  competition in synchronous open quantum systems},\ }\href@noop {} {\bibfield
  {journal} {\bibinfo  {journal} {arXiv preprint arXiv:2301.04322}\ } (\bibinfo
  {year} {2023}{\natexlab{a}})}\BibitemShut {NoStop}%
\bibitem [{\citenamefont {Murtadho}\ \emph
  {et~al.}(2023{\natexlab{b}})\citenamefont {Murtadho}, \citenamefont
  {Thingna},\ and\ \citenamefont
  {Vinjanampathy}}]{murtadho2023synchronization}%
  \BibitemOpen
  \bibfield  {author} {\bibinfo {author} {\bibfnamefont {T.}~\bibnamefont
  {Murtadho}}, \bibinfo {author} {\bibfnamefont {J.}~\bibnamefont {Thingna}},\
  and\ \bibinfo {author} {\bibfnamefont {S.}~\bibnamefont {Vinjanampathy}},\
  }\bibfield  {title} {\bibinfo {title} {Synchronization lower bounds the
  efficiency of near-degenerate thermal machines},\ }\href@noop {} {\bibfield
  {journal} {\bibinfo  {journal} {arXiv preprint arXiv:2301.04323}\ } (\bibinfo
  {year} {2023}{\natexlab{b}})}\BibitemShut {NoStop}%
\bibitem [{\citenamefont {Buca}\ \emph {et~al.}(2022)\citenamefont {Buca},
  \citenamefont {Booker},\ and\ \citenamefont {Jaksch}}]{buca2022algebraic}%
  \BibitemOpen
  \bibfield  {author} {\bibinfo {author} {\bibfnamefont {B.}~\bibnamefont
  {Buca}}, \bibinfo {author} {\bibfnamefont {C.}~\bibnamefont {Booker}},\ and\
  \bibinfo {author} {\bibfnamefont {D.}~\bibnamefont {Jaksch}},\ }\bibfield
  {title} {\bibinfo {title} {Algebraic theory of quantum synchronization and
  limit cycles under dissipation},\ }\href
  {https://scipost.org/10.21468/SciPostPhys.12.3.097} {\bibfield  {journal}
  {\bibinfo  {journal} {SciPost Phys.}\ }\textbf {\bibinfo {volume} {12}},\
  \bibinfo {pages} {097} (\bibinfo {year} {2022})}\BibitemShut {NoStop}%
\bibitem [{\citenamefont {Tindall}\ \emph {et~al.}(2020)\citenamefont
  {Tindall}, \citenamefont {Munoz}, \citenamefont {Bu{\v{c}}a},\ and\
  \citenamefont {Jaksch}}]{Tindall2020}%
  \BibitemOpen
  \bibfield  {author} {\bibinfo {author} {\bibfnamefont {J.}~\bibnamefont
  {Tindall}}, \bibinfo {author} {\bibfnamefont {C.~S.}\ \bibnamefont {Munoz}},
  \bibinfo {author} {\bibfnamefont {B.}~\bibnamefont {Bu{\v{c}}a}},\ and\
  \bibinfo {author} {\bibfnamefont {D.}~\bibnamefont {Jaksch}},\ }\bibfield
  {title} {\bibinfo {title} {Quantum synchronisation enabled by dynamical
  symmetries and dissipation},\ }\href
  {https://doi.org/10.1088/1367-2630/ab60f5} {\bibfield  {journal} {\bibinfo
  {journal} {New J. Phys.}\ }\textbf {\bibinfo {volume} {22}},\ \bibinfo
  {pages} {013026} (\bibinfo {year} {2020})}\BibitemShut {NoStop}%
\bibitem [{\citenamefont {Bu{\v{c}}a}\ \emph {et~al.}(2019)\citenamefont
  {Bu{\v{c}}a}, \citenamefont {Tindall},\ and\ \citenamefont
  {Jaksch}}]{buvca2019non}%
  \BibitemOpen
  \bibfield  {author} {\bibinfo {author} {\bibfnamefont {B.}~\bibnamefont
  {Bu{\v{c}}a}}, \bibinfo {author} {\bibfnamefont {J.}~\bibnamefont
  {Tindall}},\ and\ \bibinfo {author} {\bibfnamefont {D.}~\bibnamefont
  {Jaksch}},\ }\bibfield  {title} {\bibinfo {title} {Non-stationary coherent
  quantum many-body dynamics through dissipation},\ }\href
  {https://doi.org/10.1038/s41467-019-09757-y} {\bibfield  {journal} {\bibinfo
  {journal} {Nat. Commun.}\ }\textbf {\bibinfo {volume} {10}},\ \bibinfo
  {pages} {1730} (\bibinfo {year} {2019})}\BibitemShut {NoStop}%
\bibitem [{\citenamefont {Lidar}\ \emph {et~al.}(1998)\citenamefont {Lidar},
  \citenamefont {Chuang},\ and\ \citenamefont {Whaley}}]{lidar1998decoherence}%
  \BibitemOpen
  \bibfield  {author} {\bibinfo {author} {\bibfnamefont {D.~A.}\ \bibnamefont
  {Lidar}}, \bibinfo {author} {\bibfnamefont {I.~L.}\ \bibnamefont {Chuang}},\
  and\ \bibinfo {author} {\bibfnamefont {K.~B.}\ \bibnamefont {Whaley}},\
  }\bibfield  {title} {\bibinfo {title} {Decoherence-free subspaces for quantum
  computation},\ }\href {https://doi.org/10.1103/PhysRevLett.81.2594}
  {\bibfield  {journal} {\bibinfo  {journal} {Phys. Rev. Lett.}\ }\textbf
  {\bibinfo {volume} {81}},\ \bibinfo {pages} {2594} (\bibinfo {year}
  {1998})}\BibitemShut {NoStop}%
\bibitem [{\citenamefont {Affleck}\ \emph {et~al.}(1987)\citenamefont
  {Affleck}, \citenamefont {Kennedy}, \citenamefont {Lieb},\ and\ \citenamefont
  {Tasaki}}]{PhysRevLett.59.799}%
  \BibitemOpen
  \bibfield  {author} {\bibinfo {author} {\bibfnamefont {I.}~\bibnamefont
  {Affleck}}, \bibinfo {author} {\bibfnamefont {T.}~\bibnamefont {Kennedy}},
  \bibinfo {author} {\bibfnamefont {E.~H.}\ \bibnamefont {Lieb}},\ and\
  \bibinfo {author} {\bibfnamefont {H.}~\bibnamefont {Tasaki}},\ }\bibfield
  {title} {\bibinfo {title} {Rigorous results on valence-bond ground states in
  antiferromagnets},\ }\href {https://doi.org/10.1103/PhysRevLett.59.799}
  {\bibfield  {journal} {\bibinfo  {journal} {Phys. Rev. Lett.}\ }\textbf
  {\bibinfo {volume} {59}},\ \bibinfo {pages} {799} (\bibinfo {year}
  {1987})}\BibitemShut {NoStop}%
\bibitem [{\citenamefont {Haldane}(1983)}]{haldane1983continuum}%
  \BibitemOpen
  \bibfield  {author} {\bibinfo {author} {\bibfnamefont {F.}~\bibnamefont
  {Haldane}},\ }\bibfield  {title} {\bibinfo {title} {Continuum dynamics of the
  1-d heisenberg antiferromagnet: Identification with the o(3) nonlinear sigma
  model},\ }\href
  {https://doi.org/https://doi.org/10.1016/0375-9601(83)90631-X} {\bibfield
  {journal} {\bibinfo  {journal} {Physics Letters A}\ }\textbf {\bibinfo
  {volume} {93}},\ \bibinfo {pages} {464} (\bibinfo {year} {1983})}\BibitemShut
  {NoStop}%
\bibitem [{\citenamefont {Pollmann}\ \emph {et~al.}(2010)\citenamefont
  {Pollmann}, \citenamefont {Turner}, \citenamefont {Berg},\ and\ \citenamefont
  {Oshikawa}}]{PhysRevB.81.064439}%
  \BibitemOpen
  \bibfield  {author} {\bibinfo {author} {\bibfnamefont {F.}~\bibnamefont
  {Pollmann}}, \bibinfo {author} {\bibfnamefont {A.~M.}\ \bibnamefont
  {Turner}}, \bibinfo {author} {\bibfnamefont {E.}~\bibnamefont {Berg}},\ and\
  \bibinfo {author} {\bibfnamefont {M.}~\bibnamefont {Oshikawa}},\ }\bibfield
  {title} {\bibinfo {title} {Entanglement spectrum of a topological phase in
  one dimension},\ }\href {https://doi.org/10.1103/PhysRevB.81.064439}
  {\bibfield  {journal} {\bibinfo  {journal} {Phys. Rev. B}\ }\textbf {\bibinfo
  {volume} {81}},\ \bibinfo {pages} {064439} (\bibinfo {year}
  {2010})}\BibitemShut {NoStop}%
\bibitem [{\citenamefont {Moudgalya}\ \emph {et~al.}(2018)\citenamefont
  {Moudgalya}, \citenamefont {Rachel}, \citenamefont {Bernevig},\ and\
  \citenamefont {Regnault}}]{PhysRevB.98.235155}%
  \BibitemOpen
  \bibfield  {author} {\bibinfo {author} {\bibfnamefont {S.}~\bibnamefont
  {Moudgalya}}, \bibinfo {author} {\bibfnamefont {S.}~\bibnamefont {Rachel}},
  \bibinfo {author} {\bibfnamefont {B.~A.}\ \bibnamefont {Bernevig}},\ and\
  \bibinfo {author} {\bibfnamefont {N.}~\bibnamefont {Regnault}},\ }\bibfield
  {title} {\bibinfo {title} {Exact excited states of nonintegrable models},\
  }\href {https://doi.org/10.1103/PhysRevB.98.235155} {\bibfield  {journal}
  {\bibinfo  {journal} {Phys. Rev. B}\ }\textbf {\bibinfo {volume} {98}},\
  \bibinfo {pages} {235155} (\bibinfo {year} {2018})}\BibitemShut {NoStop}%
\bibitem [{\citenamefont {Schollw{\"o}ck}(2011)}]{schollwock2011density}%
  \BibitemOpen
  \bibfield  {author} {\bibinfo {author} {\bibfnamefont {U.}~\bibnamefont
  {Schollw{\"o}ck}},\ }\bibfield  {title} {\bibinfo {title} {The density-matrix
  renormalization group in the age of matrix product states},\ }\href
  {https://doi.org/https://doi.org/10.1016/j.aop.2010.09.012} {\bibfield
  {journal} {\bibinfo  {journal} {Ann. Phys.}\ }\textbf {\bibinfo {volume}
  {326}},\ \bibinfo {pages} {96} (\bibinfo {year} {2011})},\ \bibinfo {note}
  {january 2011 Special Issue}\BibitemShut {NoStop}%
\bibitem [{\citenamefont {Perez-Garcia}\ \emph {et~al.}(2007)\citenamefont
  {Perez-Garcia}, \citenamefont {Verstraete}, \citenamefont {Wolf},\ and\
  \citenamefont {Cirac}}]{10.5555/2011832.2011833}%
  \BibitemOpen
  \bibfield  {author} {\bibinfo {author} {\bibfnamefont {D.}~\bibnamefont
  {Perez-Garcia}}, \bibinfo {author} {\bibfnamefont {F.}~\bibnamefont
  {Verstraete}}, \bibinfo {author} {\bibfnamefont {M.~M.}\ \bibnamefont
  {Wolf}},\ and\ \bibinfo {author} {\bibfnamefont {J.~I.}\ \bibnamefont
  {Cirac}},\ }\bibfield  {title} {\bibinfo {title} {Matrix product state
  representations},\ }\href {https://dl.acm.org/doi/10.5555/2011832.2011833}
  {\bibfield  {journal} {\bibinfo  {journal} {Quantum Info. Comput.}\ }\textbf
  {\bibinfo {volume} {7}},\ \bibinfo {pages} {401} (\bibinfo {year}
  {2007})}\BibitemShut {NoStop}%
\bibitem [{\citenamefont {Or{\'u}s}(2014)}]{orus2014practical}%
  \BibitemOpen
  \bibfield  {author} {\bibinfo {author} {\bibfnamefont {R.}~\bibnamefont
  {Or{\'u}s}},\ }\bibfield  {title} {\bibinfo {title} {A practical introduction
  to tensor networks: Matrix product states and projected entangled pair
  states},\ }\href {https://doi.org/https://doi.org/10.1016/j.aop.2014.06.013}
  {\bibfield  {journal} {\bibinfo  {journal} {Ann. Phys.}\ }\textbf {\bibinfo
  {volume} {349}},\ \bibinfo {pages} {117} (\bibinfo {year}
  {2014})}\BibitemShut {NoStop}%
\bibitem [{\citenamefont {Breuer}\ and\ \citenamefont
  {Petruccione}(2002)}]{BreuerPetruccioneBook2002}%
  \BibitemOpen
  \bibfield  {author} {\bibinfo {author} {\bibfnamefont {H.~P.}\ \bibnamefont
  {Breuer}}\ and\ \bibinfo {author} {\bibfnamefont {F.}~\bibnamefont
  {Petruccione}},\ }\href@noop {} {\emph {\bibinfo {title} {The Theory of Open
  Quantum Systems}}}\ (\bibinfo  {publisher} {Oxford University Press},\
  \bibinfo {address} {Oxford},\ \bibinfo {year} {2002})\BibitemShut {NoStop}%
\bibitem [{\citenamefont {Schaller}(2014)}]{SchallerBook2014}%
  \BibitemOpen
  \bibfield  {author} {\bibinfo {author} {\bibfnamefont {G.}~\bibnamefont
  {Schaller}},\ }\href@noop {} {\emph {\bibinfo {title} {Open Quantum Systems
  Far from Equilibrium}}}\ (\bibinfo  {publisher} {Lect. Notes Phys.,
  Springer},\ \bibinfo {address} {Cham},\ \bibinfo {year} {2014})\BibitemShut
  {NoStop}%
\bibitem [{\citenamefont {Kraus}\ \emph {et~al.}(2008)\citenamefont {Kraus},
  \citenamefont {B\"uchler}, \citenamefont {Diehl}, \citenamefont {Kantian},
  \citenamefont {Micheli},\ and\ \citenamefont
  {Zoller}}]{kraus2008preparation}%
  \BibitemOpen
  \bibfield  {author} {\bibinfo {author} {\bibfnamefont {B.}~\bibnamefont
  {Kraus}}, \bibinfo {author} {\bibfnamefont {H.~P.}\ \bibnamefont
  {B\"uchler}}, \bibinfo {author} {\bibfnamefont {S.}~\bibnamefont {Diehl}},
  \bibinfo {author} {\bibfnamefont {A.}~\bibnamefont {Kantian}}, \bibinfo
  {author} {\bibfnamefont {A.}~\bibnamefont {Micheli}},\ and\ \bibinfo {author}
  {\bibfnamefont {P.}~\bibnamefont {Zoller}},\ }\bibfield  {title} {\bibinfo
  {title} {Preparation of entangled states by quantum markov processes},\
  }\href {https://doi.org/10.1103/PhysRevA.78.042307} {\bibfield  {journal}
  {\bibinfo  {journal} {Phys. Rev. A}\ }\textbf {\bibinfo {volume} {78}},\
  \bibinfo {pages} {042307} (\bibinfo {year} {2008})}\BibitemShut {NoStop}%
\bibitem [{\citenamefont {Zhou}\ \emph {et~al.}(2021)\citenamefont {Zhou},
  \citenamefont {Choi},\ and\ \citenamefont {Lukin}}]{zhou2021symmetry}%
  \BibitemOpen
  \bibfield  {author} {\bibinfo {author} {\bibfnamefont {L.}~\bibnamefont
  {Zhou}}, \bibinfo {author} {\bibfnamefont {S.}~\bibnamefont {Choi}},\ and\
  \bibinfo {author} {\bibfnamefont {M.~D.}\ \bibnamefont {Lukin}},\ }\bibfield
  {title} {\bibinfo {title} {Symmetry-protected dissipative preparation of
  matrix product states},\ }\href {https://doi.org/10.1103/PhysRevA.104.032418}
  {\bibfield  {journal} {\bibinfo  {journal} {Phys. Rev. A}\ }\textbf {\bibinfo
  {volume} {104}},\ \bibinfo {pages} {032418} (\bibinfo {year}
  {2021})}\BibitemShut {NoStop}%
\bibitem [{\citenamefont {Sch\"on}\ \emph {et~al.}(2005)\citenamefont
  {Sch\"on}, \citenamefont {Solano}, \citenamefont {Verstraete}, \citenamefont
  {Cirac},\ and\ \citenamefont {Wolf}}]{PhysRevLett.95.110503}%
  \BibitemOpen
  \bibfield  {author} {\bibinfo {author} {\bibfnamefont {C.}~\bibnamefont
  {Sch\"on}}, \bibinfo {author} {\bibfnamefont {E.}~\bibnamefont {Solano}},
  \bibinfo {author} {\bibfnamefont {F.}~\bibnamefont {Verstraete}}, \bibinfo
  {author} {\bibfnamefont {J.~I.}\ \bibnamefont {Cirac}},\ and\ \bibinfo
  {author} {\bibfnamefont {M.~M.}\ \bibnamefont {Wolf}},\ }\bibfield  {title}
  {\bibinfo {title} {Sequential generation of entangled multiqubit states},\
  }\href {https://doi.org/10.1103/PhysRevLett.95.110503} {\bibfield  {journal}
  {\bibinfo  {journal} {Phys. Rev. Lett.}\ }\textbf {\bibinfo {volume} {95}},\
  \bibinfo {pages} {110503} (\bibinfo {year} {2005})}\BibitemShut {NoStop}%
\bibitem [{\citenamefont {Huang}\ and\ \citenamefont
  {Chen}(2015)}]{PhysRevB.91.195143}%
  \BibitemOpen
  \bibfield  {author} {\bibinfo {author} {\bibfnamefont {Y.}~\bibnamefont
  {Huang}}\ and\ \bibinfo {author} {\bibfnamefont {X.}~\bibnamefont {Chen}},\
  }\bibfield  {title} {\bibinfo {title} {Quantum circuit complexity of
  one-dimensional topological phases},\ }\href
  {https://doi.org/10.1103/PhysRevB.91.195143} {\bibfield  {journal} {\bibinfo
  {journal} {Phys. Rev. B}\ }\textbf {\bibinfo {volume} {91}},\ \bibinfo
  {pages} {195143} (\bibinfo {year} {2015})}\BibitemShut {NoStop}%
\bibitem [{\citenamefont {Wei}\ \emph {et~al.}(2023)\citenamefont {Wei},
  \citenamefont {Malz},\ and\ \citenamefont
  {Cirac}}]{PhysRevResearch.5.L022037}%
  \BibitemOpen
  \bibfield  {author} {\bibinfo {author} {\bibfnamefont {Z.-Y.}\ \bibnamefont
  {Wei}}, \bibinfo {author} {\bibfnamefont {D.}~\bibnamefont {Malz}},\ and\
  \bibinfo {author} {\bibfnamefont {J.~I.}\ \bibnamefont {Cirac}},\ }\bibfield
  {title} {\bibinfo {title} {Efficient adiabatic preparation of tensor network
  states},\ }\href {https://doi.org/10.1103/PhysRevResearch.5.L022037}
  {\bibfield  {journal} {\bibinfo  {journal} {Phys. Rev. Res.}\ }\textbf
  {\bibinfo {volume} {5}},\ \bibinfo {pages} {L022037} (\bibinfo {year}
  {2023})}\BibitemShut {NoStop}%
\bibitem [{\citenamefont {Kaltenbaek}\ \emph {et~al.}(2010)\citenamefont
  {Kaltenbaek}, \citenamefont {Lavoie}, \citenamefont {Zeng}, \citenamefont
  {Bartlett},\ and\ \citenamefont {Resch}}]{kaltenbaek2010optical}%
  \BibitemOpen
  \bibfield  {author} {\bibinfo {author} {\bibfnamefont {R.}~\bibnamefont
  {Kaltenbaek}}, \bibinfo {author} {\bibfnamefont {J.}~\bibnamefont {Lavoie}},
  \bibinfo {author} {\bibfnamefont {B.}~\bibnamefont {Zeng}}, \bibinfo {author}
  {\bibfnamefont {S.~D.}\ \bibnamefont {Bartlett}},\ and\ \bibinfo {author}
  {\bibfnamefont {K.~J.}\ \bibnamefont {Resch}},\ }\bibfield  {title} {\bibinfo
  {title} {Optical one-way quantum computing with a simulated valence-bond
  solid},\ }\href@noop {} {\bibfield  {journal} {\bibinfo  {journal} {Nat.
  Phys.}\ }\textbf {\bibinfo {volume} {6}},\ \bibinfo {pages} {850} (\bibinfo
  {year} {2010})}\BibitemShut {NoStop}%
\bibitem [{\citenamefont {Murta}\ \emph {et~al.}(2023)\citenamefont {Murta},
  \citenamefont {Cruz},\ and\ \citenamefont
  {Fern\'andez-Rossier}}]{PhysRevResearch.5.013190}%
  \BibitemOpen
  \bibfield  {author} {\bibinfo {author} {\bibfnamefont {B.}~\bibnamefont
  {Murta}}, \bibinfo {author} {\bibfnamefont {P.~M.~Q.}\ \bibnamefont {Cruz}},\
  and\ \bibinfo {author} {\bibfnamefont {J.}~\bibnamefont
  {Fern\'andez-Rossier}},\ }\bibfield  {title} {\bibinfo {title} {Preparing
  valence-bond-solid states on noisy intermediate-scale quantum computers},\
  }\href {https://doi.org/10.1103/PhysRevResearch.5.013190} {\bibfield
  {journal} {\bibinfo  {journal} {Phys. Rev. Res.}\ }\textbf {\bibinfo {volume}
  {5}},\ \bibinfo {pages} {013190} (\bibinfo {year} {2023})}\BibitemShut
  {NoStop}%
\bibitem [{\citenamefont {Chen}\ \emph {et~al.}(2023)\citenamefont {Chen},
  \citenamefont {Shen}, \citenamefont {Lee},\ and\ \citenamefont
  {Yang}}]{10.21468/SciPostPhys.15.4.170}%
  \BibitemOpen
  \bibfield  {author} {\bibinfo {author} {\bibfnamefont {T.}~\bibnamefont
  {Chen}}, \bibinfo {author} {\bibfnamefont {R.}~\bibnamefont {Shen}}, \bibinfo
  {author} {\bibfnamefont {C.~H.}\ \bibnamefont {Lee}},\ and\ \bibinfo {author}
  {\bibfnamefont {B.}~\bibnamefont {Yang}},\ }\bibfield  {title} {\bibinfo
  {title} {{High-fidelity realization of the AKLT state on a NISQ-era quantum
  processor}},\ }\href {https://doi.org/10.21468/SciPostPhys.15.4.170}
  {\bibfield  {journal} {\bibinfo  {journal} {SciPost Phys.}\ }\textbf
  {\bibinfo {volume} {15}},\ \bibinfo {pages} {170} (\bibinfo {year}
  {2023})}\BibitemShut {NoStop}%
\bibitem [{\citenamefont {Smith}\ \emph {et~al.}(2023)\citenamefont {Smith},
  \citenamefont {Crane}, \citenamefont {Wiebe},\ and\ \citenamefont
  {Girvin}}]{PRXQuantum.4.020315}%
  \BibitemOpen
  \bibfield  {author} {\bibinfo {author} {\bibfnamefont {K.~C.}\ \bibnamefont
  {Smith}}, \bibinfo {author} {\bibfnamefont {E.}~\bibnamefont {Crane}},
  \bibinfo {author} {\bibfnamefont {N.}~\bibnamefont {Wiebe}},\ and\ \bibinfo
  {author} {\bibfnamefont {S.}~\bibnamefont {Girvin}},\ }\bibfield  {title}
  {\bibinfo {title} {Deterministic constant-depth preparation of the aklt state
  on a quantum processor using fusion measurements},\ }\href
  {https://doi.org/10.1103/PRXQuantum.4.020315} {\bibfield  {journal} {\bibinfo
   {journal} {PRX Quantum}\ }\textbf {\bibinfo {volume} {4}},\ \bibinfo {pages}
  {020315} (\bibinfo {year} {2023})}\BibitemShut {NoStop}%
\bibitem [{\citenamefont {Wang}\ \emph {et~al.}(2023)\citenamefont {Wang},
  \citenamefont {Snizhko}, \citenamefont {Romito}, \citenamefont {Gefen},\ and\
  \citenamefont {Murch}}]{wang2023dissipative}%
  \BibitemOpen
  \bibfield  {author} {\bibinfo {author} {\bibfnamefont {Y.}~\bibnamefont
  {Wang}}, \bibinfo {author} {\bibfnamefont {K.}~\bibnamefont {Snizhko}},
  \bibinfo {author} {\bibfnamefont {A.}~\bibnamefont {Romito}}, \bibinfo
  {author} {\bibfnamefont {Y.}~\bibnamefont {Gefen}},\ and\ \bibinfo {author}
  {\bibfnamefont {K.}~\bibnamefont {Murch}},\ }\bibfield  {title} {\bibinfo
  {title} {Dissipative preparation and stabilization of many-body quantum
  states in a superconducting qutrit array},\ }\href
  {https://doi.org/10.1103/PhysRevA.108.013712} {\bibfield  {journal} {\bibinfo
   {journal} {Phys. Rev. A}\ }\textbf {\bibinfo {volume} {108}},\ \bibinfo
  {pages} {013712} (\bibinfo {year} {2023})}\BibitemShut {NoStop}%
\bibitem [{\citenamefont {Pollmann}\ \emph {et~al.}(2012)\citenamefont
  {Pollmann}, \citenamefont {Berg}, \citenamefont {Turner},\ and\ \citenamefont
  {Oshikawa}}]{PhysRevB.85.075125}%
  \BibitemOpen
  \bibfield  {author} {\bibinfo {author} {\bibfnamefont {F.}~\bibnamefont
  {Pollmann}}, \bibinfo {author} {\bibfnamefont {E.}~\bibnamefont {Berg}},
  \bibinfo {author} {\bibfnamefont {A.~M.}\ \bibnamefont {Turner}},\ and\
  \bibinfo {author} {\bibfnamefont {M.}~\bibnamefont {Oshikawa}},\ }\bibfield
  {title} {\bibinfo {title} {Symmetry protection of topological phases in
  one-dimensional quantum spin systems},\ }\href
  {https://doi.org/10.1103/PhysRevB.85.075125} {\bibfield  {journal} {\bibinfo
  {journal} {Phys. Rev. B}\ }\textbf {\bibinfo {volume} {85}},\ \bibinfo
  {pages} {075125} (\bibinfo {year} {2012})}\BibitemShut {NoStop}%
\bibitem [{sup()}]{supp}%
  \BibitemOpen
  \href@noop {} {}\bibinfo {note} {See Supplemental Material for a discussion
  about the robustness of topological synchronization and about experimental
  implementations.}\BibitemShut {Stop}%
\bibitem [{\citenamefont {Sompet}\ \emph {et~al.}(2022)\citenamefont {Sompet},
  \citenamefont {Hirthe}, \citenamefont {Bourgund}, \citenamefont {Chalopin},
  \citenamefont {Bibo}, \citenamefont {Koepsell}, \citenamefont {Bojovi{\'c}},
  \citenamefont {Verresen}, \citenamefont {Pollmann}, \citenamefont {Salomon}
  \emph {et~al.}}]{sompet2022realizing}%
  \BibitemOpen
  \bibfield  {author} {\bibinfo {author} {\bibfnamefont {P.}~\bibnamefont
  {Sompet}}, \bibinfo {author} {\bibfnamefont {S.}~\bibnamefont {Hirthe}},
  \bibinfo {author} {\bibfnamefont {D.}~\bibnamefont {Bourgund}}, \bibinfo
  {author} {\bibfnamefont {T.}~\bibnamefont {Chalopin}}, \bibinfo {author}
  {\bibfnamefont {J.}~\bibnamefont {Bibo}}, \bibinfo {author} {\bibfnamefont
  {J.}~\bibnamefont {Koepsell}}, \bibinfo {author} {\bibfnamefont
  {P.}~\bibnamefont {Bojovi{\'c}}}, \bibinfo {author} {\bibfnamefont
  {R.}~\bibnamefont {Verresen}}, \bibinfo {author} {\bibfnamefont
  {F.}~\bibnamefont {Pollmann}}, \bibinfo {author} {\bibfnamefont
  {G.}~\bibnamefont {Salomon}}, \emph {et~al.},\ }\bibfield  {title} {\bibinfo
  {title} {Realizing the symmetry-protected haldane phase in fermi--hubbard
  ladders},\ }\href@noop {} {\bibfield  {journal} {\bibinfo  {journal}
  {Nature}\ }\textbf {\bibinfo {volume} {606}},\ \bibinfo {pages} {484}
  (\bibinfo {year} {2022})}\BibitemShut {NoStop}%
\bibitem [{\citenamefont {Cattaneo}\ \emph {et~al.}(2023)\citenamefont
  {Cattaneo}, \citenamefont {Rossi}, \citenamefont {Garc\'{\i}a-P\'erez},
  \citenamefont {Zambrini},\ and\ \citenamefont
  {Maniscalco}}]{PRXQuantum.4.010324}%
  \BibitemOpen
  \bibfield  {author} {\bibinfo {author} {\bibfnamefont {M.}~\bibnamefont
  {Cattaneo}}, \bibinfo {author} {\bibfnamefont {M.~A.}\ \bibnamefont {Rossi}},
  \bibinfo {author} {\bibfnamefont {G.}~\bibnamefont {Garc\'{\i}a-P\'erez}},
  \bibinfo {author} {\bibfnamefont {R.}~\bibnamefont {Zambrini}},\ and\
  \bibinfo {author} {\bibfnamefont {S.}~\bibnamefont {Maniscalco}},\ }\bibfield
   {title} {\bibinfo {title} {Quantum simulation of dissipative collective
  effects on noisy quantum computers},\ }\href
  {https://doi.org/10.1103/PRXQuantum.4.010324} {\bibfield  {journal} {\bibinfo
   {journal} {PRX Quantum}\ }\textbf {\bibinfo {volume} {4}},\ \bibinfo {pages}
  {010324} (\bibinfo {year} {2023})}\BibitemShut {NoStop}%
\bibitem [{\citenamefont {Norcia}\ and\ \citenamefont
  {Thompson}(2016)}]{PhysRevX.6.011025}%
  \BibitemOpen
  \bibfield  {author} {\bibinfo {author} {\bibfnamefont {M.~A.}\ \bibnamefont
  {Norcia}}\ and\ \bibinfo {author} {\bibfnamefont {J.~K.}\ \bibnamefont
  {Thompson}},\ }\bibfield  {title} {\bibinfo {title} {Cold-strontium laser in
  the superradiant crossover regime},\ }\href
  {https://doi.org/10.1103/PhysRevX.6.011025} {\bibfield  {journal} {\bibinfo
  {journal} {Phys. Rev. X}\ }\textbf {\bibinfo {volume} {6}},\ \bibinfo {pages}
  {011025} (\bibinfo {year} {2016})}\BibitemShut {NoStop}%
\bibitem [{\citenamefont {Angerer}\ \emph {et~al.}(2018)\citenamefont
  {Angerer}, \citenamefont {Streltsov}, \citenamefont {Astner}, \citenamefont
  {Putz}, \citenamefont {Sumiya}, \citenamefont {Onoda}, \citenamefont {Isoya},
  \citenamefont {Munro}, \citenamefont {Nemoto}, \citenamefont {Schmiedmayer}
  \emph {et~al.}}]{angerer2018superradiant}%
  \BibitemOpen
  \bibfield  {author} {\bibinfo {author} {\bibfnamefont {A.}~\bibnamefont
  {Angerer}}, \bibinfo {author} {\bibfnamefont {K.}~\bibnamefont {Streltsov}},
  \bibinfo {author} {\bibfnamefont {T.}~\bibnamefont {Astner}}, \bibinfo
  {author} {\bibfnamefont {S.}~\bibnamefont {Putz}}, \bibinfo {author}
  {\bibfnamefont {H.}~\bibnamefont {Sumiya}}, \bibinfo {author} {\bibfnamefont
  {S.}~\bibnamefont {Onoda}}, \bibinfo {author} {\bibfnamefont
  {J.}~\bibnamefont {Isoya}}, \bibinfo {author} {\bibfnamefont {W.~J.}\
  \bibnamefont {Munro}}, \bibinfo {author} {\bibfnamefont {K.}~\bibnamefont
  {Nemoto}}, \bibinfo {author} {\bibfnamefont {J.}~\bibnamefont
  {Schmiedmayer}}, \emph {et~al.},\ }\bibfield  {title} {\bibinfo {title}
  {Superradiant emission from colour centres in diamond},\ }\href@noop {}
  {\bibfield  {journal} {\bibinfo  {journal} {Nat. Phys.}\ }\textbf {\bibinfo
  {volume} {14}},\ \bibinfo {pages} {1168} (\bibinfo {year}
  {2018})}\BibitemShut {NoStop}%
\end{thebibliography}
\end{document}


\title{Supplemental Material to ``Topological quantum synchronization of fractionalized spins''}

\author{Christopher W. W\"achtler}
\affiliation{Department of Physics, University of California, Berkeley, California 94720, USA}
\author{Joel E. Moore}
\affiliation{Department of Physics, University of California, Berkeley, California 94720, USA}
\affiliation{Materials Sciences Division, Lawrence Berkeley National Laboratory, Berkeley, California 94720, USA}


\maketitle

\onecolumngrid
\section{Topological protection of the synchronized dynamics}

\begin{figure*}[h!]
    \centering
\includegraphics[width=0.95\textwidth]{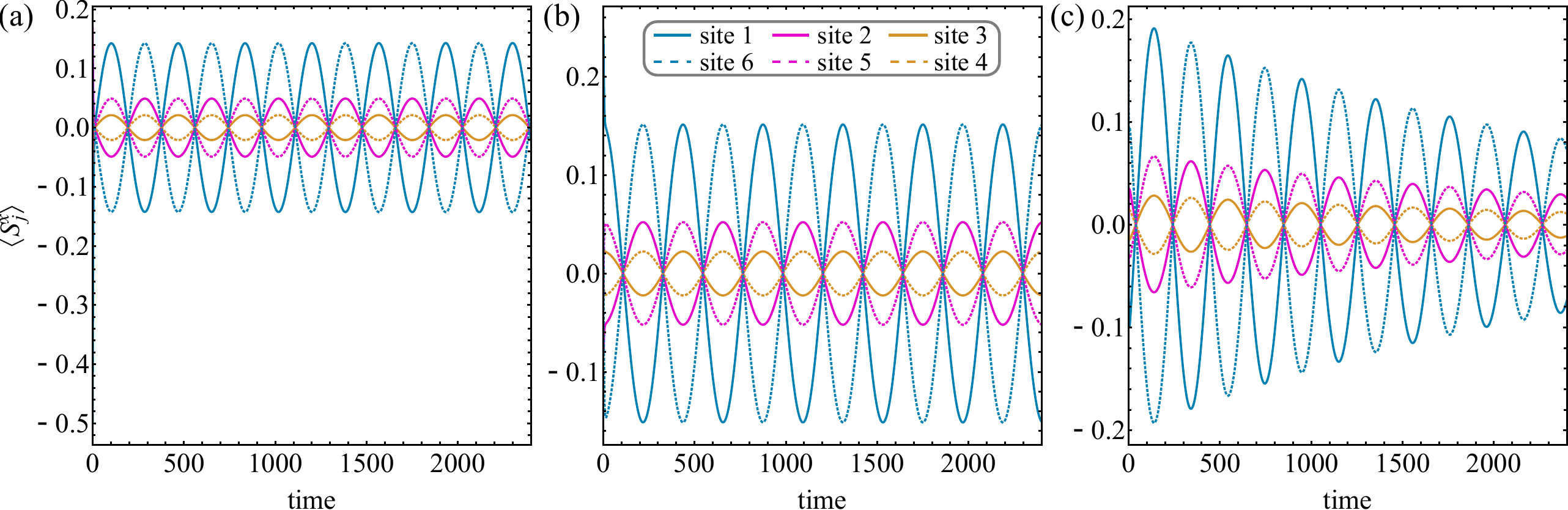}
    \caption{Robustness of topological synchronization within the ground state subspace if inversion symmetry is broken. In all three figures, the initial state is a random superposition between all four states belonging to the ground state subspace. (a) Synchronized dynamics for $J_{max} = 0.5$ and homogeneous magnetic field ($B_j=0$). (b) Additional perturbations to the magnetic field with $B_\mathrm{max} = 0.2 B$. (c) Breaking all symmetries of the AKLT model by considering an additional perpendicular magnetic field $(B_\mathrm x/N) \sum_j S^\mathrm x_j$ results in damped synchronized dynamics.}
    \label{figApp:Fig1}
\end{figure*}

The ground state manifold of the open AKLT chain with Hamiltonian
\begin{equation}
    H_\mathrm{AKLT} = \sum\limits_{j=1}^{N-1} \left[\frac{1}{2}\vec{S}_j\cdot \vec{S}_{j+1} + \frac{1}{6}\left(\vec S_{j}\cdot \vec S_{j+1}\right)^2 + \frac{1}{3}\right]
\end{equation}
belongs to the Haldane phase and is protected by three symmetries~[53,~70]: time-reversal symmetry (TRS), link inversion symmetry (lattice inversion about the center of a bond) and $\mathbb Z_2\times \mathbb Z_2$ symmetry ($\pi$ rotations about two orthogonal axes). Importantly, the AKLT ground states are protected as long as any \emph{one} of the three symmetries is preserved. In order to synchronize the non-interacting fractionalized degrees of freedom via the collective decay channel mediated through $L_\mathrm G = \sqrt{\gamma}S^-= \sqrt{\gamma} \sum_j S_j^-$, the fourfold degeneracy of the ground state manifold needs to be lifted -- at least partially -- while preserving the total magnetization ($\left[H,S^\mathrm z\right]=0$), such that different total spin sectors do not mix. A viable choice (considered in the main manuscript) is an additional magnetic field, i.e. $H_\mathrm{B} = (B/N) \sum_j S^\mathrm z_j$, which breaks TRS symmetry. However, as the other symmetries are still preserved, the system remains gapped and the ground state manifold is topologically protected. To show explicitly that a single symmetry is sufficient for synchronized dynamics, we consider in the following the Hamiltonian 
\begin{equation}
    H = \sum\limits_{j=1}^{N-1} (1-J_i)\left[\frac{1}{2}\vec{S}_j\cdot \vec{S}_{j+1} + \frac{1}{6}\left(\vec S_{j}\cdot \vec S_{j+1}\right)^2 + \frac{1}{3}\right] +\sum\limits_{j=1}^N \frac{B-B_j}{N} S^\mathrm{z}_j,
\end{equation}
where $J_i\sim \mathcal U(0,J_\mathrm{max})$ and $B_j\sim \mathcal U(0,B_\mathrm{max})$ are uniformly distributed random variables, and $0\leq J_\mathrm{max}<1$ and $0\leq B_\mathrm{max}<B$. Both the spin interactions and the inhomogeneous magnetic field break the inversion symmetry of the chain, yet the synchronized dynamics within the ground state subspace remains as shown in Figs.~\ref{figApp:Fig1}(a) and (b). However, if all three symmetries are broken by considering for example an additional perpendicular magnetic field $(B_\mathrm x/N) \sum_j S^\mathrm x_j$, synchronization is not stable in the long-time limit; see Fig.~\ref{figApp:Fig1}(c). The topological robustness of the closed chain thus directly translates to the dissipative dynamics of the fractionalized spins at the ends of the chain resulting in stable synchronization over a wide range of parameters. 

\section{Potential implementations of collective dissipation in superconducting qutrit systems}

\rev{As stated in the main manuscript, the most suitable implementation of topological quantum synchronization follows a two-step process: Firstly, preparation of the ground state manifold of the symmetry-protected topological spin-1 chain, followed by a single global dissipative process, which realizes the actual synchronization. In the following, we will discuss two approaches building on recent experimental advancements.}

\rev{The symmetry-protected Haldane phase has been realized in cold atoms using the spin-1/2 Fermi-Hubbard two-leg ladder [72]. This achievement serves as a promising starting point for observing synchronized dynamics by utilizing a cavity to implement the collective decay. However, one may also use superconducting qutrits in combination with a cavity to prepare the AKLT ground state subspace as proposed for example in Ref.~[69]. In both cases, the collective decay is achieved by operating in the so-called `bad cavity' limit, which has been realized experimentally in cold atoms [74] as well as nitrogen-vacancy centers [75].}

Assuming, that all \rev{spin-1s or} qutrits are coupled to the cavity with the same magnitude $\lambda$, the total system is given by the Hamiltonian
\begin{equation}
\label{eqApp:H_tot}
H_\mathrm{tot} = H_\mathrm{AKLT} + \lambda \sum\limits_j \left(S_j^- a^\dagger + S_j^+ a\right) + \Omega a^\dagger a, 
\end{equation}
with bosonic operators $a^\dagger$ and $a$ and cavity frequency $\Omega$. Additionally, the cavity is subject to dissipation giving rise to the dynamics 
\begin{equation}
\label{eqApp:ME}
\dot \varrho(t) = -\mathrm i\left[H_\mathrm{tot},\varrho(t)\right] + \mathcal D\left[L_\mathrm{cav}\right]\varrho(t),
\end{equation}
where $L_\mathrm{cav}= \sqrt{\Gamma}a$. In the \rev{bad} cavity limit \rev{$\Gamma >\lambda \gg \kappa$, where $\kappa$ denotes the decay rate of individual atoms, we can neglect local dissipation. Moreover,} we can adiabatically eliminate the cavity mode and assume it remains in the ground state. The Heisenberg equation of motion for the cavity mode $a$ is then given by
\begin{equation}
\dot a = -\mathrm i\lambda \left(\sum\limits_j S_j^-\right) - \left(\mathrm i \Omega + \frac{\Gamma}{2}\right)a \approx  -\mathrm i\lambda \left(\sum\limits_j S_j^-\right) - \frac{\Gamma}{2} a .
\end{equation}
In the \rev{bad} cavity limit we assume $\dot a=0$ since the presence of the coupled qutrits does not alter the cavity amplitude, which remains unchanged. The resulting expression for the cavity mode can be resubstituted into Eqs.~(\ref{eqApp:H_tot}) and (\ref{eqApp:ME}), which yields
\begin{equation}
\dot \varrho(t) = -\mathrm i\left[H_\mathrm{AKLT},\varrho(t)\right] + \frac{4\lambda^2}{\Gamma}\left(S^- \varrho(t) S^+ - \frac{1}{2}\left\{S^+ S^-, \varrho(t)\right\}\right) = -\mathrm i\left[H_\mathrm{AKLT},\varrho(t)\right] + \mathcal D\left[L_\mathrm G\right]\varrho(t)
\end{equation}
where $L_\mathrm G = \sqrt{\gamma} S^-$ with $\gamma = 4\lambda^2/ \Gamma$, and  we have neglected the cavity term because we assume that the cavity remains in the ground state, i.e. $\Omega a^\dagger a\ket{0}=0$. \rev{Note that the collective decay with rate $\gamma$ occurs on time scales much faster than individual decay processes in the bad cavity limit because we assumed $\Gamma >\lambda \gg \kappa$.}

\begin{figure*}
    \centering
\includegraphics[width=0.25\textwidth]{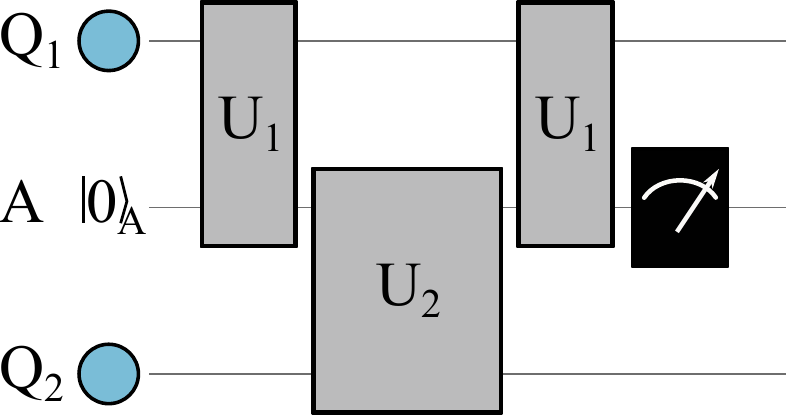}
    \caption{Circuit diagram for a single time step of the implementation of collective decay for a two-qutrit system via measurement of an ancilla system. The circuit presented here needs to applied repeatedly and ensemble averaged in order to correspond to the collective decay channel $L_\mathrm G$. The generalization to many qutrits is straight forward. }
    \label{figApp:Fig2}
\end{figure*}

If one prepares the ground state subspace of the AKLT model in terms of digital quantum simulation as for example shown recently in Ref.~[68], one can use a quantum circuit in order to effectively implement \rev{also} the global dissipation via repeatedly measuring \rev{(or replacing) an ancilla system. An example of a recent experimental implementation of collective decay in qubits can be found in Ref.~[73]}. In the following we discuss the implementation for collective decay of two qutrits explicitly but its generalization to larger chains is straightforward. The circuit that needs to be implemented is shown in Fig. \ref{figApp:Fig2}, where the system consists of two qutrits (1 and 2) and an ancilla (A). We initialize the ancilla in the ground state $\ket{0}_\mathrm{A}$. The unitary to be implement corresponds to $U_\mathrm{tot} = \exp(-\mathrm i \sqrt{\delta t \lambda} H_\mathrm A)$, where 
\begin{equation}
H_\mathrm A = \sum\limits_{j=1}^2 \left(S_j^+ \ket{0}\bra{1}_\mathrm A + S_j^- \ket{1}\bra{0}_\mathrm A\right) = H_1+H_2. 
\end{equation}
We can approximate the total unitary $U_\mathrm{tot}$ using a second order Trotter-Suzuki formula as 
\begin{equation}
U_\mathrm{tot}(\Delta t) = e^{-\mathrm i \frac{\sqrt{\Delta t \lambda}}{4}H_1} e^{-\mathrm i \frac{\sqrt{\Delta t \lambda}}{2}H_2}e^{-\mathrm i \frac{\sqrt{\Delta t \lambda}}{4}H_1} +\mathcal O(\Delta t^2) = U_1(\Delta t/4)U_2(\Delta t/2)U_1(\Delta t/4) +\mathcal O(\Delta t^2). 
\end{equation}
Note that we do not need to implement the AKLT Hamiltonian as we assume we are already in the ground state subspace. After the unitary evolution with $U_\mathrm{tot}(\Delta t)$  we measure the ancilla which yields the post-measurement state $\ket{\psi_{m}(\Delta t)}$ for the two qutrits depending on the measurement outcome $m={0,1}$. These are given by
\begin{equation}
\ket{\psi_1(\Delta t)} = -\mathrm i\frac{S^-}{\sqrt{\braket{S^+ S^-}}}\ket{\psi(t=0)} +\mathcal O(\Delta t^{3/2}),  \qquad \ket{\psi_0(\Delta t)} = \left[1-\frac{\Delta t \lambda}{8}\left(S^+ S^- + \braket{S^+ S^-}\right)\right]\ket{\psi(t=0)} +\mathcal O(\Delta t^{3/2}),  
\end{equation}
where $\ket{\psi(t=0)}$ is the initial state of the two qutrits before the unitary and the measurement. After measurement, the ancilla is reset to state $\ket{0}_\mathrm A$ \rev{(or replaced by a new qubit)}. This realizes a stochastic Schrödinger equation for the two qutrits with a single stochastic step given by 
\begin{equation}
\ket{d\psi} = \ket{\psi(t+\Delta t)} -\ket{\psi(t)} = \left[\left(-\mathrm i\frac{S^-}{\sqrt{\braket{S^+ S^-}}}\ket{\psi}-1\right)dN - \frac{\Delta t \lambda}{8}\left(S^+ S^- + \braket{S^+ S^-}\right)\right]\ket{\psi(t)},
\end{equation}
where $dN$ is a Poison increment with $dN^2 = dN$ and $\mathbb E\left[dN\right] = (\lambda \Delta t /4) \braket{S^+ S^-}$. Hence, repeatedly evolving the qutrit-ancilla system and measuring the state of the ancilla results in a collective dissipative process. Upon ensemble averaging, this precisely implements the collective dissipator $L_\mathrm G = \sqrt{\gamma}\sum_j S^-_j$ with $\gamma = \lambda \Delta t/4$.